\documentclass[%
groupedaddress,
preprintnumbers, 
amsmath, amssymb,
aps, prl,
 twocolumn
 ]{revtex4-2}
 
\usepackage{xcolor}
\usepackage{graphicx}
\usepackage{dcolumn}
\usepackage{bm}
\usepackage[colorlinks]{hyperref}
\usepackage[mathlines]{lineno}

\graphicspath{{img/}}

\begin{document}

\title{Formation of Multiple BGK-like Structures in the Time-asymptotic State of Collisionless Vlasov-Poisson Plasmas}  

\author{Hugo A. Carril} \email{hcarril@udec.cl}
\author{Jorge A. Gidi} \email{jorgegidi@udec.cl}
\author{Roberto E. Navarro} \email{roberto.navarro@udec.cl}
\author{Jaime A. Araneda}  \email{jaraneda@udec.cl}
\affiliation{Departamento de F\'isica, Facultad de Ciencias F\'isicas
  y Matem\'aticas, Universidad de Concepci\'on, Concepci\'on Chile}

\begin{abstract}
  The time-asymptotic state of a finite-amplitude perturbation in a
  collisionless and Maxwellian plasma is typically represented as a
  steady state of two non-linearly superposed, counter-propagating
  Bernstein-Greene-Kruskal (BGK) modes.  Using high-resolution
  Vlasov-Poisson simulations, we show that the plasma evolves
  self-consistently into a time-asymptotic state of multiple
  vortex-like structures that gradually fill the phase-space and
  reduce filamentation.  This occurs without the need for external
  forcing or the presence of an energetic plasma population. This finding suggests that the time-asymptotic regime of the plasma is
  rather akin to a non-linear superposition of multiple BGK-like modes
  associated with nearly constant phase-speed waves.  The electric
  field and the space-averaged particle distribution function exhibit
  a power-law broad spectrum, which is consistent with an energy
  cascade towards smaller scales in both position and velocity spaces.
\end{abstract}
   
\maketitle

\section{Introduction}
Finite-amplitude spatial perturbations or external forcing in a collisionless 
plasma can lead to undamped, solitary, and 
non-linear states in  which resonant particles  
become trapped in potential wells. This is observed as localized 
vortices or holes in the phase-space of particles, which can last for long periods of
time~\cite{PhysRevLett.79.2815,PhysRevLett.81.5137,raghunathan2013nonlinear,trivedi2017driven,hutchinson2017electron}. 
In particular, this situation is found during the 
non-linear Landau damping of Langmuir waves, in which an initial
electrostatic perturbation on an unmagnetized electron plasma leads to
a saturated state where electrons are trapped by the electrostatic
potential around the phase-speed of the perturbed wave
mode~\cite{PhysRevLett.79.2815,ivanov2004wave,PhysRevE.68.026406,xu2017effects}.
Similar phenomena have been found in observational research as
trapping of electrons along the magnetic field lines in near-Earth
environments~\cite{tong2018simultaneous,lotekar2020multisatellite,steinvall2019multispacecraft,mozer2018direct,norgren2015slow,graham2016electrostatic,holmes2018electron};
in experimental research dedicated to Landau damping and particle
trapping~\cite{PhysRevLett.92.245003,PhysRevLett.29.914,saeki1979formation};
in electron beam injection into low-$\beta$ plasma
columns~\cite{lefebvre2010laboratory}; and in simulations of
collisionless electron plasmas with
Maxwellian~\cite{rupp2015density,PhysRevLett.79.2815,trivedi2016chirp,PhysRevE.71.017402}
and non-Maxwellian distribution
functions~\cite{valentini2005nonlinear,raghunathan2013nonlinear,trivedi2017driven,lopez2017landau}.

Phase-space vortices in electrostatic plasmas are exact solutions of the 
one-dimensional Vlasov-Poisson system of equations known as
Bernstein-Greene-Kruskal (BGK)
states~\cite{bernstein1957exact,schamel1979theory,schamel2012cnoidal,schamel2015particle}.
For a BGK state to be stable the distribution function of the trapped particles 
must not be monotonically decreasing~\cite{pankavich2014instability},
and the background distribution supporting 
particle trapping must have less than
three maxima~\cite{manfredi2000stability}. These 
conditions allow the propagation of stable
phase-space holes for long times in plasmas with 
single-maximum distributions, such as Maxwellian 
distributions~\cite{manfredi2000stability}.

The long-time stability of one-dimensional BGKs are affected 
by numerical diffusive effects caused by the finite resolution 
of the position and velocity grids, modifying 
the phase-space structures observed in
Vlasov-Poisson
simulations~\cite{morse1969one,galeotti2005asymptotic,califano2006vlasov}.
Also, multiple one-dimensional
phase-space holes are unstable to the growth of 
sidebands, which disrupt the state by 
merging the holes when the wavelength of the 
perturbations is smaller than the length of the 
spatial domain~\cite{shoucri2017formation,brunetti2000asymptotic,pandey2021trapped}. Indeed, it has been 
observed that multiple neighboring BGK
structures evolve 
toward a time-asymptotic state consisting of a
single stable hole in long-term 
Vlasov-Poisson simulations~\cite{ghizzo1988stability}.

The formation
of multiple regions of particle trapping in
the distribution function can be
approximately represented as a 
non-linear superposition of individual 
BGK modes, where the electrostatic 
potentials are linearly superposed
and the distribution function is built in a 
piece-wise way~\cite{PhysRevLett.70.3732,
PhysRevE.50.1465}. This approximation is 
supported by Vlasov 
simulations~\cite{demeio1990numerical,demeio2001numerical,fijalkow2005superposition,brunetti2000asymptotic} and by 
theoretical works~\cite{PhysRevE.68.026406,PhysRevLett.81.5137,lancellotti1999time,lancellotti2009nonlinear}. 
However, 
this non-linear superposition does
not capture all structures formed in the
distribution function~\cite{demeio1990numerical,demeio2001numerical}.

This work addresses a long-standing academic question about the long-time
evolution of the non-linear Landau damping problem. To date, there are two known paradigms 
describing the time-asymptotic 
state of one-dimensional 
Maxwellian electron 
plasmas undergoing non-linear Landau damping. One of them states that
the time-asymptotic state corresponds to a non-linear superposition of two individual counter-propagating BGK
states~\cite{PhysRevLett.79.2815}. In this case, the phase space develops stable vortical structures after Landau damping stops~\cite{rupp2015density}, with the space-averaged distributions exhibiting characteristic plateaus at velocities around the BGK's phase-velocities~\cite{brunetti2000asymptotic,PhysRevLett.79.2815,galeotti2005asymptotic,raghunathan2013nonlinear}.
The second paradigm 
states that the time-asymptotic state is 
akin to a non-linear superposition of
multiple BGK 
states~\cite{lancellotti1999time,PhysRevLett.81.5137}.

Our results suggest that the physical picture of the time-asymptotic
state of collisionless Vlasov-Poisson plasmas is far richer than the
standard view of two stable BGK modes only, thus rather akin to the
framework of multiple BGK-like
states~\cite{PhysRevLett.81.5137}. Indeed, it is shown that multiple
small phase-space holes occur in a wide range of velocities below the
Langmuir phase-speed, $|v|<v_L$. Each hole is associated with electron
plasma waves (EPWs) whose phase speeds are similar to the propagation speed of the corresponding holes. The electric field reaches a
time-asymptotic state characterized by a power-law energy spectrum,
indicating an energy cascade process. In addition, the space-averaged
distribution function exhibits a piece-wise power-law velocity spectrum, showing a reduction in phase-space filamentation, suggesting that the
development of small-scale electron
holes is self-similar and intermittent. 

\section{Vlasov-Poisson Numerical Simulations}
The propagation of high-frequency oscillations in 
collisionless and unmagnetized plasmas is studied in terms 
of the one-dimensional Vlasov-Poisson system of equations: 
\begin{align}
    \frac{\partial f}{\partial t} + v\frac{\partial f}{\partial x} + \frac{e}{m} \frac{\partial\phi}{\partial x} \frac{\partial f}{\partial v} = 0\,, \label{eq:vlasov}\\
    \frac{\partial^2 \phi}{\partial x^2} = 4\pi e  \left(\int_{-\infty}^\infty dv\, f - n_0\right)\,, \label{eq:poisson}
\end{align}
where $f=f(x,v,t)$ is the electron phase-space 
distribution function; $\phi=\phi(x,t)$ is the 
electrostatic potential; and $-e$, $m$, and $n_0$ 
are the charge, mass, and density of electrons, 
respectively. The ions are considered as a 
motionless neutralizing background with 
density $n_i=n_0$, and periodic boundary conditions are assumed in $x$.

To solve Eqs.~\eqref{eq:vlasov} and~\eqref{eq:poisson}, an FFT
pseudospectral code was written in the Julia programming
language~\cite{Julia}. A semi-Lagrangian splitting scheme is
employed~\cite{sonnendrucker1999semi}, with spectral interpolation in the
Fourier transformed phase-space. The time advance is performed using an
optimized second-order symplectic integrator~\cite{mclachlan1992}.
Spectral interpolation introduces an exponential error that decreases with the number of nodes~\cite{fornberg1994review,boyd2001chebyshev}, so that less error is achieved for the same number of nodes compared to, for example, 
polynomial-spline interpolation.  A Gaussian-like filter of the form
$e^{-36(\nu/\nu_{\rm max})^{36}}$~\cite{parker2015fourier} is used to
smooth out abrupt variations of the velocity distribution function at
small velocity scales, where $\nu$ is the conjugate variable of $v$,
and $\nu_{\rm max}$ is the maximum allowed value of $\nu$. Although
not shown here, the Vlasov code has been successfully tested on several problems, e.g.~\textcite{PhysRevLett.79.2815,raghunathan2013nonlinear,galeotti2005asymptotic,rupp2015density,califano2006vlasov}, among others.

The initially single-mode perturbed 
electron phase-space distribution is
\begin{align}
  f(x,v,0) =  f_0(v) \left[1 + 
               \varepsilon\cos(k_1 x) \right]\,,
\end{align} 
where $\varepsilon$ and $k_1$ are the amplitude and 
wavenumber of the perturbation, respectively, and 
$f_0(v)$ the unperturbed Maxwellian distribution 
function, 
\begin{align}
  f_0(v) = \frac{n_0}{\sqrt{2\pi} v_{\rm th}}
           \exp\left( -\frac{v^2}{2 v_{\rm th}^2} \right)\,,
\end{align}
where $v_{\rm{}th}=\sqrt{k_B T/m}$ is the electron 
thermal speed, $k_B$ the Boltzmann constant, 
and $T$ the electron temperature.

Simulations are performed with a box-length $L = 2\pi/k_1$, and
velocity domain $|v|/v_{\rm th} < 7.7$; phase-space is discretized
into $N_x=2^{11}=2048$ and $N_v = 2^{17}=131072$
grid points in the space and velocity
components, respectively; the time-step is $\omega_p\,{\Delta}t=0.05$,
with a maximum time simulation run of $\omega_pt_{\rm max} = 14000$,
where $\omega_p = \sqrt{4{\pi}e^2n_0/m}$ is the electron plasma
frequency; and perturbation parameters are $\varepsilon=0.04$ and
$k_1v_{\rm{}th}/\omega_p=0.4$.
The effects of changing the velocity grid resolution $2^{11}\leq N_v\leq 2^{17}$ on the formation of phase-space holes are shown in later sections.

\section{Non-linear Landau damping}
\begin{figure*}[ht!]
  \centering
  \includegraphics[width=0.75\linewidth]{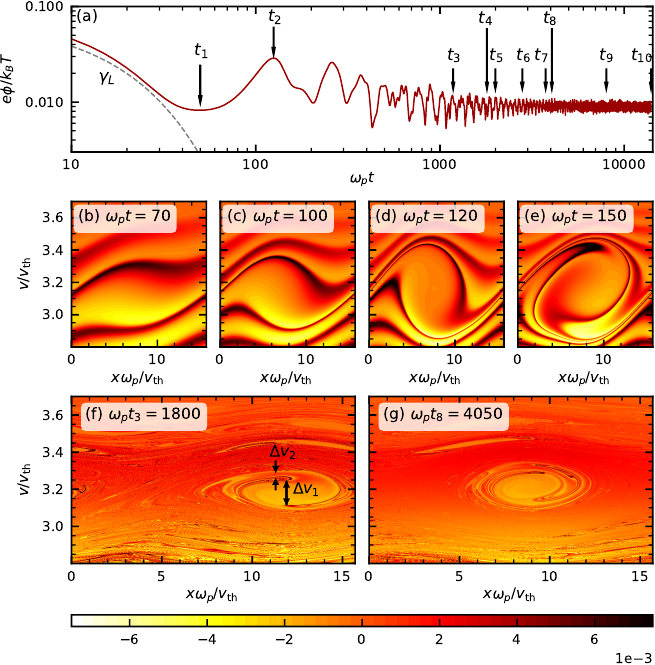}
  \caption{(a) Log-log time evolution of the 
  envelope of the first Fourier mode of the 
  electrostatic potential. The dashed gray line 
  corresponds to the linear Landau damping prediction. 
  Several times of interest are marked. 
  (b--e) Electron phase-space snapshots during the formation of the Langmuir electron-hole near $v = v_L$ at times around $t=t_2$, and (f--g) during the quasi-stationary state $t\geq t_3$. The colorbar represents the fluctuations $\delta\!f$ (in units of $n_0/v_\text{th}$) of the distribution function with respect to the unperturbed Maxwellian $f_0$.}
  \label{fig:energy}
\end{figure*}

Figure~\ref{fig:energy}(a) shows the time evolution of 
the first Fourier mode of the electrostatic potential 
in a long-time simulation. Initially, two
counter-propagating Langmuir waves are 
excited with frequencies 
$\omega_L/\omega_p\simeq\pm 1.285$ and 
phase-speeds $|v_L/v_{\rm{}th}|\simeq3.213$. 
The energy decreases 
monotonously with damping-rate 
$\gamma_L/\omega_p\simeq-0.06608$, which is consistent with 
linear Landau theory up to a time $\omega_pt<30$. 
Then, Landau damping is arrested at $\omega_pt_1
\simeq50$ and the amplitude of the
perturbed mode grows until its
saturation at a time $\omega_pt_2\simeq125$. 
After $t>t_2$ and until the end of the simulation, 
the electrostatic potential oscillates 
irregularly 
around a finite amplitude $e\phi^\ast/k_BT\simeq0.01$,
suggesting that the plasma has achieved a 
time-asymptotic state~\cite{PhysRevLett.81.5137,ivanov2004wave}.

Trapping of electrons is observed several time-steps after saturation
$t_{\rm trap}>t_2$. Indeed, Figs.~\ref{fig:energy}(b--e) show the
formation of a vortex centered at $v=v_L$ that closes completely at
$\omega_p t_{\rm trap}=150$. Due to the symmetry of the distribution
function, another vortex is also formed at $v=-v_L$. These structures
(hereafter referred to as Langmuir vortices) correspond to highly localized
deficits or holes in the electron phase-space density that rotate due
to resonant interaction with the self-consistent electric potential~\cite{hutchinson2017electron}.

Inside the vortex, inward spiral arms are formed due to the nearly
parabolic electrostatic potential at the center of the trapping
region. This sets a nearly constant bouncing frequency for the trapped
electrons~\cite{o1971nonlinear,Hara2015bunching,brunner2014kinetic,Dodin2013negativemass},
where electrons closer to the center tend to exhibit coherent
motion. Combined with the motion of particles with different velocities,
this creates a region of bunched electrons that introduces
short-wavelength fluctuations in the electrostatic potential. Toward the borders of the trapping region, the bouncing frequency for electrons varies depending on their
location and the motion is less
coherent~\cite{brunner2014kinetic}. These structures
represent strongly trapped electrons undergoing Langmuir wave
resonance.

The observed dynamics in Figs.~\ref{fig:energy}(b--g) is
also consistent with the development of the bunching instability shown
by \textcite{Hara2015bunching}.  Once the instability has saturated,
filamentation spreads particles within the hole. The process
replicates itself several times during the simulation and becomes
quasi-stationary from $\omega_pt \simeq 1300$.
Figures~\ref{fig:energy}(f--g) show that the Langmuir vortices and the spiraling
arms are persistent until the end of the simulation.

Similar structures have been observed in electron beam-plasma
configurations exhibiting instability, in 
electron beams interacting with a
trapping wave~\cite{o1972nonlinear,bouchoule1976generation,firpo2001long,carlevaro2015nonlinear}, 
and in the context of particle trapping 
generated by an external chirp-drive
as {\it shark}-like structures within the 
regions of
trapping~\cite{trivedi2016chirp,trivedi2017driven}. 
Studies on the trajectories of beam-electrons in 
a trapping potential show that particles 
may be classified into a {\it sea} of particles
with chaotic trajectories uniformly
distributed, and groups of bunched electrons
oscillating coherently~\cite{tennyson1994self,firpo2001long,farina2004large} with
stable trajectories around the 
structures~\cite{adam1981time}.

\section{Formation of small-scale phase-space vortices}
\begin{figure*}[t!]
  \centering
  \includegraphics[width=\linewidth]{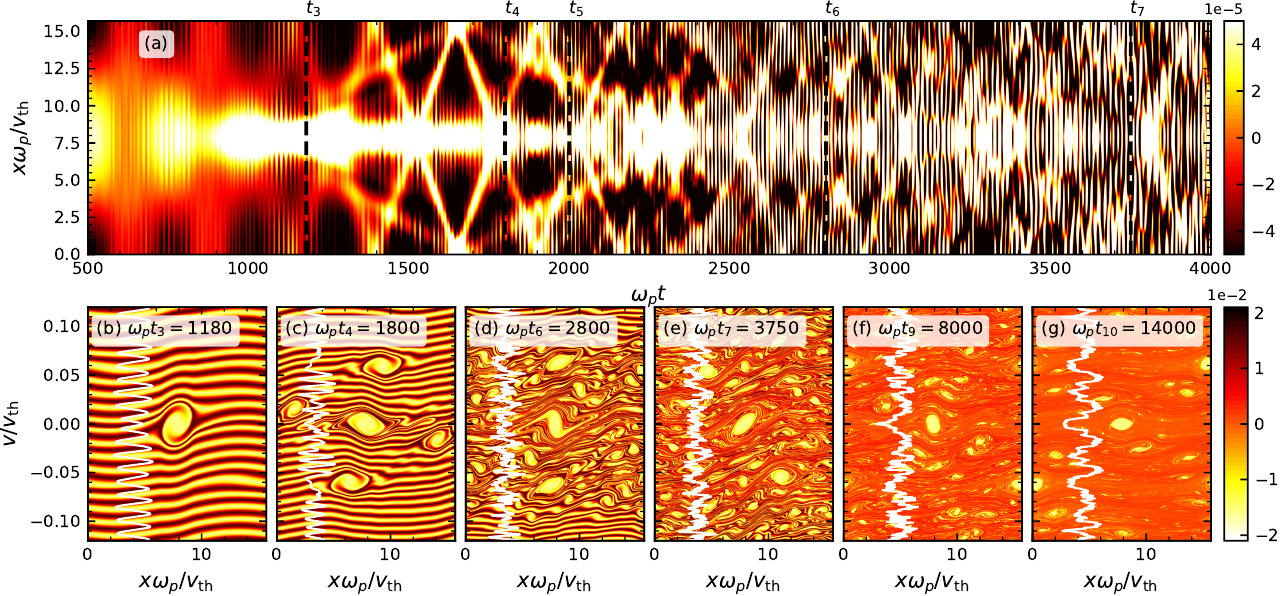}
  \caption{(a) Electrostatic potential $\phi=\phi(x,t)$ filtered to
    low frequencies $|\omega/\omega_p|<0.6$. (b--g) Fluctuations
    $\delta f = \langle{f} - f_0$, in units of $n_0/v_{\rm th}$, of
    the electron phase-space distribution for velocities around $v=0$
    and for $0<x<L$. Solid white lines are the spatially averaged
    fluctuations of the distribution function $\delta f(v) = \int dx \delta\! f(x,v)$.}
  \label{fig:potential_i}
\end{figure*}

Figure~\ref{fig:potential_i}(a) shows the electrostatic potential
$\phi=\phi(x,t)$ filtered to low frequencies $|\omega/\omega_p|<0.6$.
The superposition of counter-propagating Langmuir waves of the same
amplitude and phase speed results in a standing wave with localized
positive potential at $x=L/2$ with enhanced amplitude after
$\omega_pt>1000$ that seems to be persistent until the end of the
simulation. Then, after $\omega_pt_3 = 1180$ two pairs of solitary and
localized electrostatic waves start to propagate with velocities
$|v_1/v_{\rm{}th}|\simeq0.06$ and
$|v_2/v_{\rm{}th}|\simeq0.014$. After $t>t_5$, various modes start to grow, propagating with different but constant phase speeds.

Figure~\ref{fig:potential_i}(b) shows that at $t=t_3$ a completely closed vortex centered at $x=L/2$ and $v=0$ has formed, as expected from the symmetry in the velocity space and the beating of the waves propagating in the opposite directions. Then,
disturbances in the filamentary structure of the velocity distribution around $\pm v_1$ and $\pm v_2$
evolve into completely closed vortices at $\omega_pt_4=1800$, as shown
in Fig.~\ref{fig:potential_i}(c). The position-averaged distribution
function $\delta f = \langle f\rangle_x - f_0$, is shown as white lines
in Fig.~\ref{fig:potential_i}(b--g). In particular, at $t=t_4$ this quantity exhibits deviations from the regular oscillations seen at $t=t_3$
specially at $v=0$, $v_1$, and $v_2$. This is a signal that trapping is occurring there, and that these vortices may also be characterized as BGK-like
structures~\cite{hou2015excitation,tennyson1994self,rupp2015density}.

Note that the velocities of the vortices in
Fig.~\ref{fig:potential_i}(c) are approximately equal to
$|v_1|\simeq\Delta{v_1}/2$ and $|v_2|\simeq \Delta{v_2}/2$, where
$\Delta{v_1}/v_{\rm{}th}\simeq0.12$ and
$\Delta v_2/v_{\rm{}th}\simeq0.03$ are the velocity gaps between the
spiral arms rotating inside the Langmuir vortices shown in
Fig.~\ref{fig:energy}(f). Although not shown here, as far as we have
checked, this pattern also occurs for different values of the
perturbation amplitude $\varepsilon$. This suggests that the vortices
with velocities $v_1$ and $v_2$ result from non-linear coupling
between the spiral arms inside the Langmuir holes.

At times between $\omega_pt_6=2800$ and $\omega_pt_7=3750$,
Figs.~\ref{fig:potential_i}(d--e) show that vortices of different
scale sizes start to fill the electron phase space. This resembles the
description for phase-space turbulence in electrostatic
plasmas~\cite{dupree1982theory,dupree1983growth,berman1985simulation,schamel2017nonlinear}. At
longer times $t>t_7$, Figs.~\ref{fig:potential_i}(f--g) shows that some of the vortices have disappeared, possibly by coalescence with other holes. The hole at $v=0$ is still present at the end of the
simulation run, while the holes at $v_1$ and $v_2$ look indistinguishable from the remaining holes.

\section{Electron density $\omega$-$k$ spectrum}
\begin{figure*}[t!]
  \centering
  \includegraphics[width=\linewidth]{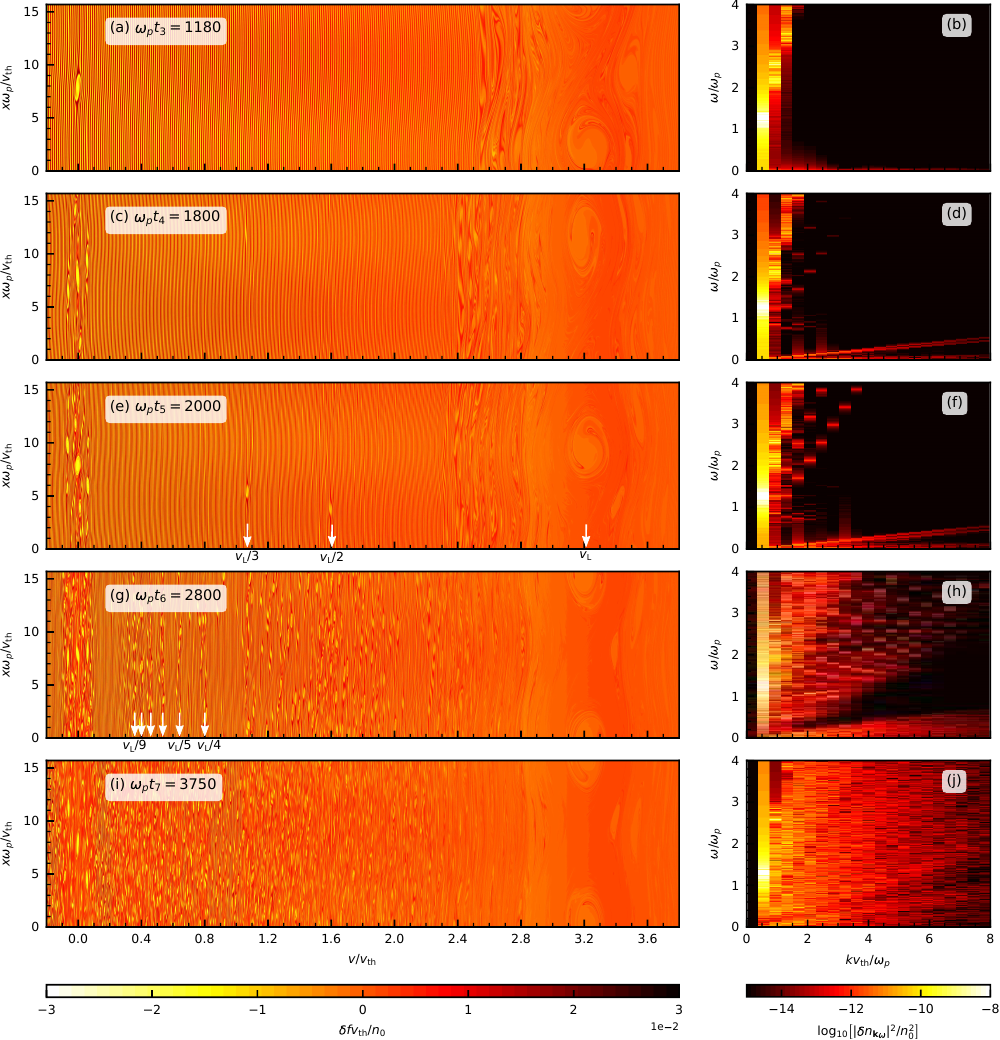}
  \caption{(left) Electron phase-space fluctuations
    $\delta f = f - f_0$, normalized to $n_0/v_{th}$; and (right)
    $\omega$-$k$ spectrum of the normalized electron density
    $\delta n_{k\omega}/n_0$ calculated in time windows defined by
    $t_i-600/\omega_p <t<t_i$, at the times $t_3$ through $t_7$
    indicated in Fig.~\ref{fig:energy}. The arrows in panels (e) and (g) represent the
    velocities $v_L/n$ where some of the vortices are formed, with $n=\{1,...,9\}$.
  }
  \label{fig:spectrum_ii}
\end{figure*}

Figure~\ref{fig:spectrum_ii}(left) shows snapshots of the full
phase-space electron density at several times up to
$\omega_pt=3750$. In addition to the vortices discussed before in
Figs.~\ref{fig:energy} and \ref{fig:potential_i}, another pair of
counter-propagating holes are visible at
$v = \pm1.07v_{\rm{}th}\simeq \pm v_L/3$, and $v = \pm1.61v_{\rm{}th}\simeq \pm v_L/2$, as shown in Fig.~\ref{fig:spectrum_ii}(e) at $\omega_p t_5=2000$.  At the times between $t_5$ and $t_6$, vortices of different sizes and localized perturbations in the distribution function appear at several specific velocities within the
range $0.2\leq|v/v_{\rm{}th}|\leq2.1$, some of them satisfying
$|v|\simeq{v_L/n}$ with $2\leq n\leq 9$. Then, at times
$\omega_p t>2800$, Figs~\ref{fig:potential_i}(g) and (i) show that
localized perturbations eventually evolve to form small-scale holes at
velocities greater than the thermal speed, eventually filling the whole phase-space.
Although not shown here, most of these vortices are eventually dissipated by coalescence or numerical dissipation. However, we still observe the presence of small-scale vortices scattered in phase-space at the end of the simulation run at $\omega_pt=14000$, just as described for Fig.~\ref{fig:potential_i}(g).

Figure~\ref{fig:spectrum_ii}(right) shows snapshots of the electron
density $\omega$-$k$ spectrum at the same times as in
Fig.~\ref{fig:spectrum_ii}(left).  The Fourier spectrum in time is
calculated with moving windows so that $t_i-600/\omega_p<t<t_i$.  At
$t=t_3$, a signal with frequency $\omega=0$ and spectrum between
$0<v_{\rm{}th}k/\omega_p<4$ can be seen in
Fig.~\ref{fig:spectrum_ii}(b), which is consistent with the formation
of a standing wave, and the formation of a stationary hole in
phase-space as described before. Then, Fig.~\ref{fig:spectrum_ii}(d)
shows that two branches of EPWs are excited at a time $t=t_4$. Their
phase-speeds $|v/v_{\rm{}th}|\simeq0.015$ and
$|v/v_{\rm{}th}| \simeq0.06$ are approximately equal to the velocities
$v_1$ and $v_2$ of the four phase-space vortices appearing around
$v=0$ in Fig.~\ref{fig:potential_i}(b).

At $t=t_5$, Fig.~\ref{fig:spectrum_ii}(f) shows the excitation of EPW
modes of phase-speeds $|v/v_{\rm{}th}|\simeq1.07\simeq v_L/3$, and
$|v/v_{\rm{}th}| \simeq1.59 \simeq v_L/2$, which is consistent with
the holes shown in Fig.~\ref{fig:spectrum_ii}(e). At longer times,
figure~\ref{fig:spectrum_ii}(h) shows that the density spectrum is
populated by several discrete EPW branches with phase-speeds starting
from $v/v_{\rm th}\simeq1$ towards $v=0$. This is correlated with the
velocity of several of the holes in phase-space with velocities
$0.23\leq|v/v_{\rm th}|\leq 1.2$ and other structures with velocities
$1.2 \leq|v/v_{\rm th}| \leq 2.8$ shown in panel
Fig.~\ref{fig:spectrum_ii}(g). At $t=t_7$, the density spectrum in
Fig.~\ref{fig:spectrum_ii}(j) is populated by a large number of EPW
branches, resembling the spectrum of spontaneous thermal fluctuations
of electrostatic waves~\cite{Yoon2007}.

\section{Quasi-stationary state}
\begin{figure}[t!]
  \centering
  \includegraphics[width=\linewidth]{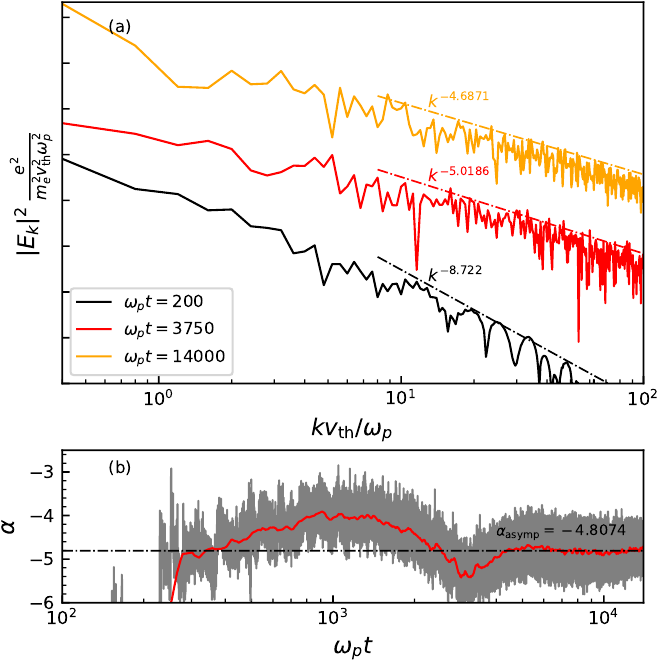}
  \caption{(a) Snapshots of the electric field 
  energy spectrum at $\omega_pt = 200$, $3750$, and $14000$. Curves are displaced vertically for 
  illustrative purposes. These spectra follow a power-law of the form $|E_k|^2\propto k^\alpha$. (b) Time evolution of the index $\alpha$. The gray line is the instantaneous value of $\alpha$, and the red line is its smoothed pattern. }
  \label{fig:electric_field_energy_spectrum}
\end{figure}

Figure~\ref{fig:electric_field_energy_spectrum}(a) shows snapshots of
the electric field energy spectrum at three different times, all of
them exhibiting a power-law wavenumber spectrum
$|E_k|^2 \propto
k^{\alpha}$. Figure~\ref{fig:electric_field_energy_spectrum}(b) shows
the evolution of the $\alpha$ index. This index varies between
$-6<\alpha<-3$ when the phase-space is being filled by small-scale
electron holes at times $400 \lesssim\omega_pt \lesssim 3000$. In particular, when the central holes are being
formed, a peak $\alpha_{\rm peak}\simeq -4$ is found around
$\omega_pt \sim 1200$. Then, $\alpha$ decreases when further
structures are developed in the range $|v| \leq v_L$. At
$\omega_pt \sim 4000$, $\alpha$ sets into a quasi-stationary state
reaching the average value $\alpha_{\rm asymp} = -4.8074$. In this
state, the phase-space is mostly filled with small-scale electron
holes. The development of this power-law spectrum suggests that the
electric field fluctuations exhibits self-similarity, meaning that the
fluctuations at different scales differ only by a scaling factor. This
process involves non-linear interactions between the nearest modes of
the electric field. Although not shown here, the electric field
auto-correlation also shows a power-law wavenumber spectrum,
suggesting that there is an energy cascade process. Due to Gauss's law, the charge density also exhibits a power-law spectrum, so the spatial electron density fluctuations also exhibit self-similarity.

\begin{figure}[t!]
  \centering
  \includegraphics[width=\linewidth]{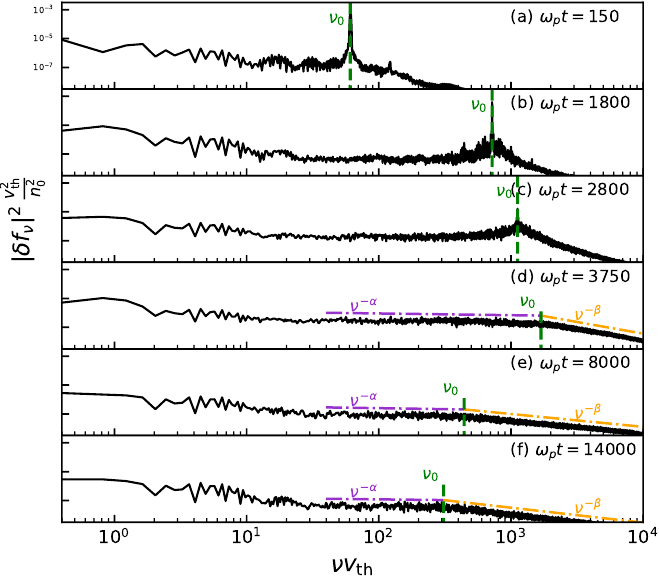}
  \caption{Fourier transform of the space-averaged distribution
    function versus the conjugate of the velocity, $\nu$. Snapshots at six different times are shown. In panels (a--c), filamentation is clearly observed as a peak in the $|\delta f_\nu|^2$ spectrum at $\nu=\nu_0$. After $\omega_p t\geq 3750$, panels (d--f), filamentation has smoothed, and $|\delta f_\nu|^2$ exhibits a break-point at $\nu=\nu_0$ separating a nearly flat spectrum $|\delta f_\nu|^2\propto \nu^{-\alpha}$ with $\alpha\simeq 0$ for $\nu<\nu_0$, and a steeper spectrum $|\delta f_\nu|^2\propto \nu^{-\beta}$ with $\beta>\alpha$ for $\nu<\nu_0$. }
  \label{fig:distribution_spectrum_filamentation}
\end{figure}

Figure~\ref{fig:distribution_spectrum_filamentation} shows snapshots
of the velocity spectrum of the space-averaged phase-space
distribution fluctuations $\delta f = \langle f \rangle_x - f_0$. A sharp peak at $\nu=\nu_0$, which corresponds to the velocity scale of the filamentation in the distribution function, can be seen in
Figs.~\ref{fig:distribution_spectrum_filamentation}(a--c). Since $\nu_0$ increases as time passes, filamentation evolves towards smaller scales in velocity space. In addition, the amplitude of the peak in $\nu_0$ decreases with time, which means that filamentation is being dissipated.

After $\omega_pt = 3750$,
Figs.~\ref{fig:distribution_spectrum_filamentation}(d--f) show that
the spectrum peak amplitude is reduced and becomes comparable to other
fluctuations, meaning that filamentation is indistinguishable from
other fluctuations in the phase-space distribution function. Then, a
piece-wise power-law spectrum of $|\delta f_\nu|^2$ is formed,
exhibiting a break-point at $\nu=\nu_0$ which separates a nearly flat
spectrum $|\delta f_\nu|^2\propto \nu^{-\alpha}$ with $\alpha\approx0$
for $\nu<\nu_0$, and a steeper spectrum
$|\delta f_\nu|^2\propto \nu^{-\beta}$ with $\beta>\alpha$ at smaller
scales $\nu>\nu_0$. At this stage, the distribution function becomes
filled with small-scale electron holes, suggesting that the power
contained by the filamentation is transferred to the development of
small-scale electron-holes.

This finding complements the physical picture usually
provided for filamentation, where the process is described to continue
to unbounded small
scales~\cite{klimas2018absence,vinas2018flux,besse2003semi,einkemmer2018low,klimas1994splitting}.
That is, the filamentation evolves toward small scales, but it is
self-consistently reduced by the development of small-scale vortices
across the distribution. Possibly, small-scale holes could not have
been developed if fluctuations were not transported via
filamentation. Thus, integration schemes using aggressive velocity
filters~\cite{klimas1994splitting,klimas2018absence,vinas2018flux}
hide part of the physics shown in this manuscript by eliminating the
filamentation process early in the simulations.

\begin{figure}[t!]
  \centering
  \includegraphics[width=\linewidth]{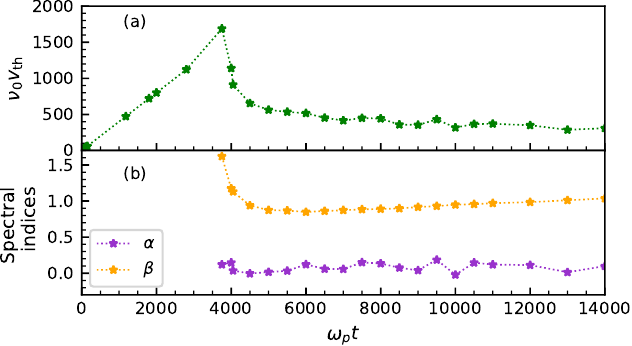}
  \caption{(a) Time evolution of the velocity scale $\nu_0$ shown in
    Fig.~\ref{fig:distribution_spectrum_filamentation}. For
    $\omega_pt<3750$, $\nu_0$ represents the velocity scale of the
    filamentation in phase-space. For $\omega_pt>3750$, $\nu_0$
    represents the break-point in $\nu$ above which the velocity
    spectra in $|\delta f_\nu|^2$ steepens. (b) Time evolution of the
    power-law spectral indices of
    $|\delta f_\nu|^2\propto \nu^{-\alpha}$ for $\nu<\nu_0$, and
    $|\delta f_\nu|^2\propto \nu^{-\beta}$ for $\nu>\nu_0$.}
  \label{fig:distribution_spectrum_vortices}
\end{figure}

Figure~\ref{fig:distribution_spectrum_vortices}(a) plots the time
evolution of $\nu_0$. Before $\omega_p t<3750$, the velocity scale of
the filamentation $\nu_0$ grows in time with a nearly constant rate
$\nu_0\approx 0.4 \omega_p t / v_{\rm th}$. Then, filamentation
becomes indistinguishable from other fluctuations at
$\omega_pt=3750$. At this point, $\nu_0$ can be identified as the
break point in the velocity spectrum of $|\delta f_\nu|^2$. It can be
seen that $\nu_0$ decreases monotonously and asymptotically approaching to $\nu_0 v_{th} \simeq 300$ for
$\omega_pt>3750$. Figure~\ref{fig:distribution_spectrum_vortices}(b)
shows the evolution power-law indices $\alpha$ and $\beta$ of the
velocity spectrum $|\delta f_\nu|^2\propto \nu^{-\alpha}$ for
$\nu<\nu_0$, and $|\delta f_\nu|^2\propto \nu^{-\beta}$ for
$\nu>\nu_0$. At all times, $\alpha\approx 0$ meaning that a nearly
equal amount of kinetic energy per mode is stored at velocity scales
$\nu<\nu_0$. On the other hand, $\beta$ quickly decreases between $3750<\omega_pt<6000$, then reaches a quasi-stationary state around $\beta\simeq1$.

The results from Figs.~\ref{fig:spectrum_ii}--\ref{fig:distribution_spectrum_vortices}
suggest that the phase-space distribution becomes self-similar,
i.e. that the small-scale electron holes formed across the
distribution follow scaling laws both in the space and velocity
domains. As time advances, the large-scale structures in the velocity
domain inherit this feature, suggesting that there are non-linear
interaction processes occurring toward structures wider in velocity
space. This self-similarity feature is probably of local character,
meaning that intermittency is happening in the development of the
small-scale electron holes.

\section{Effects of the grid resolution}
\begin{figure*}[t!]
  \centering
  \includegraphics[width=\linewidth]{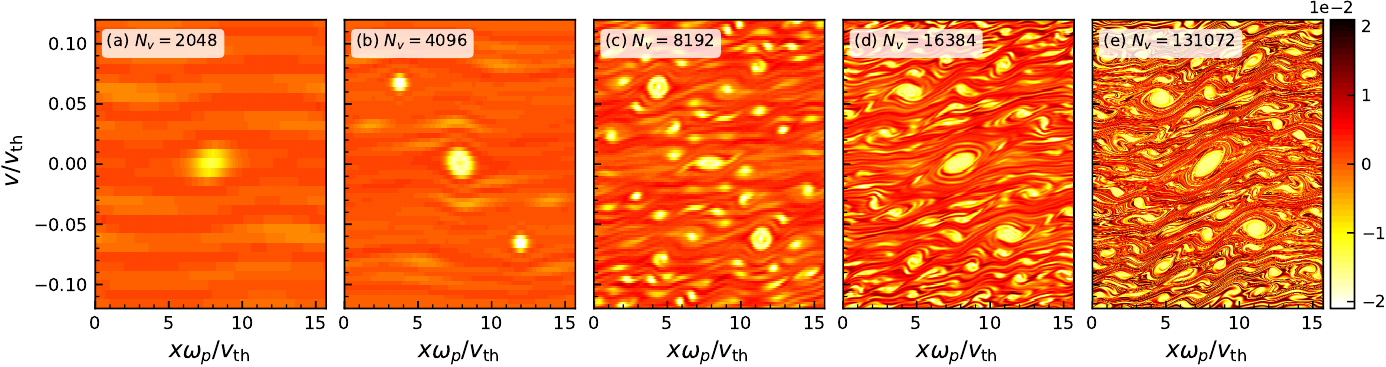}
  \caption{Electron phase-space snapshots at 
  $\omega_pt_7 = 3750$ for low velocities near $v=0$. Different 
  numbers of nodes $N_v$
  are used: (a) $N_v = 2048$, (b) $4096$,
  (c) $8192$, (d) $16384$, and
  (e) $131072$.}
  \label{fig:langmuir_different_Nv_3750_center}
\end{figure*}

The vortices observed in phase space depend critically on the grid resolution used to numerically resolve the plasma evolution, in particular on the number of nodes $N_v$ in velocity
space. Figure~\ref{fig:langmuir_different_Nv_3750_center} shows snapshots of the phase-space near $v=0$ at $\omega_pt=3750$ for five
different simulation runs with different velocity-grid resolutions $N_v$.  Although the central hole is formed in all cases,
Figs.~\ref{fig:langmuir_different_Nv_3750_center}(a--c) show that
there is a lack of small-scale structures for values of $N_v<8192$,
such as the spiraling arms or even the smaller holes formation, compared to cases with higher $N_v$ values as in
Figs.~\ref{fig:langmuir_different_Nv_3750_center}(d--e).  Simulations with small $N_v$ cannot resolve fine structures in phase space because the size of the structures reaches the grid size earlier in the simulations. In this situation, some structures do not form or their development is numerically diffusive.

\begin{figure}[t!]
  \centering
  \includegraphics[width=\linewidth]{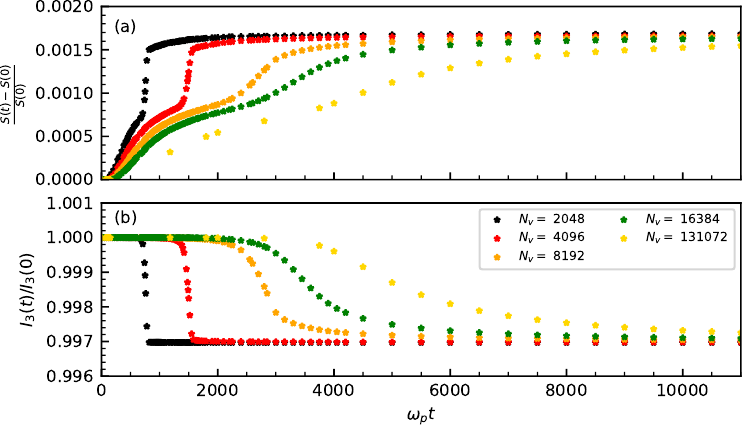}
  \caption{Time evolution of (a) the entropy and (b) the invariant $I_3$ of
  simulations run with different resolution in velocity domain. }
  \label{fig:invariant_quantities}
\end{figure}

The variations due to changes in the resolution can be quantified with
the numerical evolution of invariants of the Vlasov equation. In
Fig.~\ref{fig:invariant_quantities}, we show the time evolution of the
invariant $I_3$ and the entropy $S$, defined by
\begin{align}
  \label{eq:invariant_i3}
  I_3 &=  \int dx\,dv\, f^3\,,\\
   \label{eq:entropy}
   S &= -\int dx\,dv\, f \ln f \,.
\end{align}

Although not shown, the relative error of the total energy in all
cases is of the order of $10^{-7}$, and no significant changes are
observed by changing the number of nodes in the physical space,
$N_x$. Figure~\ref{fig:invariant_quantities} shows that a low
resolution in velocity space leads to an abrupt increment of the
entropy, and an abrupt decrement of the $I_3$ invariant at earlier
stages of the simulation. This is because fluctuations of the
distribution function $f$ reach scales close to the velocity-grid
resolution. Consequently, numerical dissipation is introduced, and
structures of the distribution cannot be solved
appropriately. Therefore, the presented results corroborate the
conclusion of \textcite{galeotti2005asymptotic}, in which the
time-asymptotic state is dependent on the dissipation caused by the
numerical grid resolution. Even if the value of $N_v$ is much higher,
deviations from invariance in both $S$ and $I_3$ are found. In this
regard, Fig.~\ref{fig:invariant_quantities} suggests that deviations
cannot be completely eliminated, but at least delayed by increasing
the velocity grid resolution $N_v$.

\section{Effects of the perturbation amplitude}
\begin{figure}[t!]
  \centering
  \includegraphics[width=\linewidth]{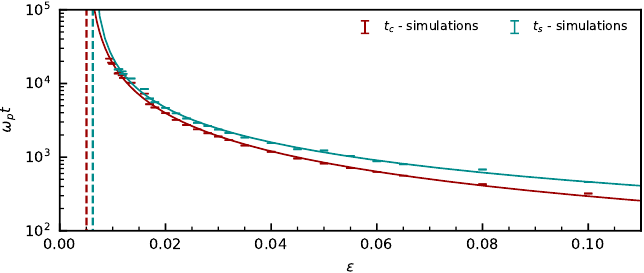}
  \caption{Times $t_c$ and $t_s$ in which the 
  central hole at $v=0$ and the secondary holes around 
  it appear, respectively, as a function of the 
  perturbation amplitude $\varepsilon$. Solid lines 
  show a power-law fit of the corresponding data. 
  Dashed lines indicate the critical amplitudes 
  for each curve. }
  \label{fig:critical}
\end{figure}

Lastly, the effects of the perturbation amplitude $\varepsilon$ on the
formation of the first holes around the center of the distribution
function are studied. Figure~\ref{fig:critical} shows the times $t_c$,
in red, at which the central hole ($v=0$) is fully formed as in
Fig.~\ref{fig:potential_i}(b) for different values of $\varepsilon$.
Similarly, the first secondary holes that appear after the central
hole as inf Fig.~\ref{fig:potential_i}(c) are also tracked. The
corresponding times $t_s$ are shown with blue error bars in
Fig.~\ref{fig:critical}. The holes are considered as fully formed when
their spiraling arms with $\delta f > 0$ coincide for the first time
in position space. The error bar size for $t_c$ and $t_s$ depends on
the sampling ratio of the distribution function around the time of
their formation. Simulations consider $N_v = 8192$ for
$\varepsilon > 0.04$, $N_v = 16384$ for $0.03 < \varepsilon < 0.04$,
$N_v = 32768$ for $ 0.0166 < \varepsilon < 0.03$, $N_v = 65536$ for
$0.016< \varepsilon <0.0166$, $N_v = 131072$ for
$0.0135< \varepsilon <0.016$, and $N_v = 262144$ for
$0.0094<\varepsilon <0.0135$.  The number of nodes in velocity space
is increased because the vortices of interest are formed at longer
times for lower values of the perturbation amplitude $\varepsilon$. In addition, finer and finer structures develop in velocity space, eventually reaching the grid resolution. Thus, $N_v$ is increased to avoid early numerical dissipation of these vortices. We checked that different values of $N_v$ do not alter the time in which the central electron holes start to form. Here, all the simulations use
$N_x = 512$, without significant differences from other values of $N_x$.

Figure~\ref{fig:critical} shows that both times $t_c$ and $t_s$
satisfy a power-law dependence on the perturbation amplitude as
$t_{i} \propto (\varepsilon - \varepsilon_i)^{\mu_i}$, where
$\varepsilon_i$ represents a critical amplitude above which slow holes
with $v\approx0$ can appear.  For the central vortex, it is found that
$\varepsilon_c=0.005\pm0.002$ and $\mu_c=-1.4\pm0.1$. For the
secondary holes growing from the central hole,
$\varepsilon_s=0.007\pm0.002$ and $\mu_s=-1.3\pm0.1$. This suggests
that the formation of the central and secondary holes in the
distribution function may be identified as a second-order transition
process~\cite{stanley1973introduction}. That is, depending on the initial amplitude perturbation $\varepsilon$, the phase-space state
switches from a phase with pure filamentation to another phase where the filamentation is reduced to form small-scale vortices in phase-space.

\section{Conclusions}
The problem of the time-asymptotic state of a collisionless, initially
perturbed Maxwellian distributed electron plasma with immobile ions
has been revisited. High-resolution Vlasov-Poisson simulations show
the generation of small-scale vortices in phase-space other than the
already known vortices formed by Landau resonance. Spiraling arms are
formed within the Langmuir vortices, corresponding to strongly trapped electrons
bouncing within the trapping potential. Then, phase-space vortices are
formed near velocities $v\approx 0$ of the phase-space electron
distribution function.

As the plasma evolves, the phase-space of electrons is successively
filled with small-scale holes below the Langmuir phase-speed, with
some holes satisfying a simple relationship $v = v_L/n$, where $v_L$
is the Langmuir phase-speed, and $n$ is an integer number. In this
process, the energy stored by the filamentation is transferred toward
ever smaller scales in velocity space. This process reduces the
filamentation while small-electron holes are formed across the
distribution. This indicates that the filamentation has scale bounds
in velocity space, rather than continuing indefinitely toward smaller
scales~\cite{klimas2018absence,klimas1994splitting,vinas2018flux,besse2003semi,
  einkemmer2018low}. The small-scale electron holes seem to persist
for long times, and we associate them as small-scale BGK-like
states. Accordingly, the physical picture of the time-asymptotic
behavior of the perturbed Maxwellian electron plasma is far more
complex than the typical description of a state consisting of only two counter-propagating
electron holes with velocities near the Langmuir phase-speed~\cite{PhysRevLett.79.2815,raghunathan2013nonlinear,brunetti2000asymptotic,demeio1990numerical,demeio2001numerical,pandey2021trapped}.
This is more consistent with the notion of multiple small-amplitude
BGK-like
states~\cite{PhysRevLett.81.5137,lancellotti1999time,PhysRevE.50.1465}.

It is worth noticing that a state consisting of multiple BGK-like
structures has already been reported in Maxwellian electron plasmas
induced by chirping of an external electrostatic
forcing~\cite{trivedi2017driven}. There, the reduction of the
frequency displaces the resonance zone toward the center of the
distribution function, yielding persisting phase-space holes. However,
the state of multiple BGK-like structures reported in this manuscript
is created self-consistently from the initial perturbation of a single
mode, without the need for an external
forcing~\cite{trivedi2017driven} nor the presence of an energetic population of particles, such as electron
beams~\cite{o1971nonlinear,adam1981time,tennyson1994self,Hara2015bunching}.

The small-amplitude electron holes are probably grown due to
non-linear wave-wave coupling between spiraling arms inside the
Langmuir holes, satisfying a beating condition
$\omega_k=\omega_{k-\ell}+\omega_\ell$. Then, a wave coupling cascade
between electron plasma waves (EPWs) starts to fill the $\omega$-$k$ power
spectrum. Whenever a hole appears, a branch of EPWs with phase-speed
equal to the velocity of the holes is excited. After the distribution
function is filled with small-scale electron holes, the energy
spectrum of the electric field and of its auto-correlations are found to obey a power-law
dependency on the wavenumber. This suggests that non-linear interactions occur
between the closest modes so that an energy cascade process is in
play.

The velocity spectrum of the space-averaged distribution function
exhibits a piece-wise power law, with a flat spectrum at large
velocity scales, a steepened spectrum for small velocity scales, and a break-point that separates both scales. Thus, fluctuations in the velocity
distribution function exhibit self-similarity. As time advances,
small-scale electron holes start to localize, suggesting the presence
of intermittency, and rendering the self-similarity as local. As the filamentation becomes negligible and small-scale electron holes are formed
across the phase-space, it is observed that the slope of the electric
field energy spectrum acquires a quasi-steady behavior.

Also, the time at which the slowest phase-space electron holes are
formed follow a power-law function of the initial amplitude
perturbation $\varepsilon$, suggesting that a process similar to a
critical phenomenon is occurring. Thus, there exists a critical
amplitude perturbation $\varepsilon_i$, below which the phase-space
filamentation is reduced indefinitely until it reaches the numerical
cell size, and above which the filamentation is smoothed to form
small-scale electron vortices in phase-space.

The findings of this research correspond to an extension of the common
interpretation of the time-asymptotic state of non-linear Landau
damping in a one-dimensional setting consisting of only two
counter-propagating electron holes. An interesting question is if this
time-asymptotic state could be attained in 2D or 3D simulations. For example, electron holes are observed to propagate stably in multidimensional magnetized plasma simulations~\cite{wu2010transverse,hutchinson2018transverse,hutchinson2019transverse}.

This study also highlights the need for high-resolution
simulations and algorithms with low numerical dissipation. This is
particularly important when addressing time-asymptotic states, because
small-scale structures that seem to be relevant in the evolution of
the system may not be appropriately resolved. The use of lower
resolution, at least in the velocity dependency, leads to an abrupt
increment in theoretically invariant quantities (e.g. the entropy and the cubic $I_3$ invariant) earlier in the
simulation. Thus, simulations with low resolution in the velocity space may dissipate the formation of structures in phase space, preventing the plasma from reaching the state observed in our
simulations.

The dissipation is intrinsically caused by the numerical scheme chosen
to solve the Vlasov-Poisson system. For example,  polynomial-splines of
second~\cite{raghunathan2013nonlinear} and third
order~\cite{PhysRevLett.79.2815,galeotti2005asymptotic} have an error
of order $O(N_v^{-n-1})$, where $n$ is the order of the
interpolation~\cite{mangeney2002numerical,colella1984piecewise,arber2002critical}. The
integration scheme used in this research has spectral interpolation
in the velocity dependency, which leads to an error of exponential
order decreasing with
$N_v$~\cite{fornberg1994review,boyd2001chebyshev}, thus reducing significantly the dissipation by numerical interpolation.

Lastly, high-order modes must be retained when addressing the velocity
dependency in Fourier space. Accordingly, it is advised to consider
large integer $n$ when using velocity filters of the form
$e^{-2n(\nu/\nu_{\rm max})^{2n}}$~\cite{Cheng1976,parker2015fourier}.
This allows the conservation of the $2n-1$ moments of the distribution
function~\cite{klimas1987method,klimas1994splitting,klimas2017simulation,klimas2018absence},
including the density and flux of particles. Values of $n$ close to
unity introduce strong damping in high-velocity modes, and although
they may allow the conservation of the lower moments of the
distribution function, they rule out the small-scale vortices observed
in this research.  In this regard, \textcite{Cheng1976} advised not to
use too large values of $n$, because their numerical experiments
indicate that the distribution function acquires negative values with
increasing $n$. However, we used a large resolution in velocity space
and the problems found by \textcite{Cheng1976} do not occur.

\section{Acknowledgements}
\begin{acknowledgments}
This work has been financially supported by ANID, Chile
through the projects FONDECyT No.~11180947 and 1191351 (R.E.N.), and
No.~1161700 (J.A.A.); CONICyT-PAI No.~79170095 (R.E.N); and the postgraduate fellowships No. 22182152 and 21202472 (H.A.C.), and No. 22182344 and 21202616 (J.A.G).
\end{acknowledgments}


\begin{thebibliography}{82}%
\makeatletter
\providecommand \@ifxundefined [1]{%
 \@ifx{#1\undefined}
}%
\providecommand \@ifnum [1]{%
 \ifnum #1\expandafter \@firstoftwo
 \else \expandafter \@secondoftwo
 \fi
}%
\providecommand \@ifx [1]{%
 \ifx #1\expandafter \@firstoftwo
 \else \expandafter \@secondoftwo
 \fi
}%
\providecommand \natexlab [1]{#1}%
\providecommand \enquote  [1]{``#1''}%
\providecommand \bibnamefont  [1]{#1}%
\providecommand \bibfnamefont [1]{#1}%
\providecommand \citenamefont [1]{#1}%
\providecommand \href@noop [0]{\@secondoftwo}%
\providecommand \href [0]{\begingroup \@sanitize@url \@href}%
\providecommand \@href[1]{\@@startlink{#1}\@@href}%
\providecommand \@@href[1]{\endgroup#1\@@endlink}%
\providecommand \@sanitize@url [0]{\catcode `\\12\catcode `\$12\catcode
  `\&12\catcode `\#12\catcode `\^12\catcode `\_12\catcode `\%12\relax}%
\providecommand \@@startlink[1]{}%
\providecommand \@@endlink[0]{}%
\providecommand \url  [0]{\begingroup\@sanitize@url \@url }%
\providecommand \@url [1]{\endgroup\@href {#1}{\urlprefix }}%
\providecommand \urlprefix  [0]{URL }%
\providecommand \Eprint [0]{\href }%
\providecommand \doibase [0]{https://doi.org/}%
\providecommand \selectlanguage [0]{\@gobble}%
\providecommand \bibinfo  [0]{\@secondoftwo}%
\providecommand \bibfield  [0]{\@secondoftwo}%
\providecommand \translation [1]{[#1]}%
\providecommand \BibitemOpen [0]{}%
\providecommand \bibitemStop [0]{}%
\providecommand \bibitemNoStop [0]{.\EOS\space}%
\providecommand \EOS [0]{\spacefactor3000\relax}%
\providecommand \BibitemShut  [1]{\csname bibitem#1\endcsname}%
\let\auto@bib@innerbib\@empty
\bibitem [{\citenamefont {Manfredi}(1997)}]{PhysRevLett.79.2815}%
  \BibitemOpen
  \bibfield  {author} {\bibinfo {author} {\bibfnamefont {G.}~\bibnamefont
  {Manfredi}},\ }\href {https://doi.org/10.1103/PhysRevLett.79.2815} {\bibfield
   {journal} {\bibinfo  {journal} {Phys. Rev. Lett.}\ }\textbf {\bibinfo
  {volume} {79}},\ \bibinfo {pages} {2815} (\bibinfo {year}
  {1997})}\BibitemShut {NoStop}%
\bibitem [{\citenamefont {Lancellotti}\ and\ \citenamefont
  {Dorning}(1998)}]{PhysRevLett.81.5137}%
  \BibitemOpen
  \bibfield  {author} {\bibinfo {author} {\bibfnamefont {C.}~\bibnamefont
  {Lancellotti}}\ and\ \bibinfo {author} {\bibfnamefont {J.~J.}\ \bibnamefont
  {Dorning}},\ }\href {https://doi.org/10.1103/PhysRevLett.81.5137} {\bibfield
  {journal} {\bibinfo  {journal} {Phys. Rev. Lett.}\ }\textbf {\bibinfo
  {volume} {81}},\ \bibinfo {pages} {5137} (\bibinfo {year}
  {1998})}\BibitemShut {NoStop}%
\bibitem [{\citenamefont {Raghunathan}\ and\ \citenamefont
  {Ganesh}(2013)}]{raghunathan2013nonlinear}%
  \BibitemOpen
  \bibfield  {author} {\bibinfo {author} {\bibfnamefont {M.}~\bibnamefont
  {Raghunathan}}\ and\ \bibinfo {author} {\bibfnamefont {R.}~\bibnamefont
  {Ganesh}},\ }\href {https://doi.org/10.1063/1.4794320} {\bibfield  {journal}
  {\bibinfo  {journal} {Phys. Plasmas}\ }\textbf {\bibinfo {volume} {20}},\
  \bibinfo {pages} {032106} (\bibinfo {year} {2013})}\BibitemShut {NoStop}%
\bibitem [{\citenamefont {Trivedi}\ and\ \citenamefont
  {Ganesh}(2017)}]{trivedi2017driven}%
  \BibitemOpen
  \bibfield  {author} {\bibinfo {author} {\bibfnamefont {P.}~\bibnamefont
  {Trivedi}}\ and\ \bibinfo {author} {\bibfnamefont {R.}~\bibnamefont
  {Ganesh}},\ }\href {https://doi.org/10.1063/1.4978560} {\bibfield  {journal}
  {\bibinfo  {journal} {Phys. Plasmas}\ }\textbf {\bibinfo {volume} {24}},\
  \bibinfo {pages} {032107} (\bibinfo {year} {2017})}\BibitemShut {NoStop}%
\bibitem [{\citenamefont {Hutchinson}(2017)}]{hutchinson2017electron}%
  \BibitemOpen
  \bibfield  {author} {\bibinfo {author} {\bibfnamefont {I.}~\bibnamefont
  {Hutchinson}},\ }\href {https://doi.org/10.1063/1.4976854} {\bibfield
  {journal} {\bibinfo  {journal} {Phys. Plasmas}\ }\textbf {\bibinfo {volume}
  {24}},\ \bibinfo {pages} {055601} (\bibinfo {year} {2017})}\BibitemShut
  {NoStop}%
\bibitem [{\citenamefont {Ivanov}\ \emph {et~al.}(2004)\citenamefont {Ivanov},
  \citenamefont {Cairns},\ and\ \citenamefont {Robinson}}]{ivanov2004wave}%
  \BibitemOpen
  \bibfield  {author} {\bibinfo {author} {\bibfnamefont {A.}~\bibnamefont
  {Ivanov}}, \bibinfo {author} {\bibfnamefont {I.~H.}\ \bibnamefont {Cairns}},\
  and\ \bibinfo {author} {\bibfnamefont {P.}~\bibnamefont {Robinson}},\ }\href
  {https://doi.org/10.1063/1.1785789} {\bibfield  {journal} {\bibinfo
  {journal} {Phys. Plasmas}\ }\textbf {\bibinfo {volume} {11}},\ \bibinfo
  {pages} {4649} (\bibinfo {year} {2004})}\BibitemShut {NoStop}%
\bibitem [{\citenamefont {Lancellotti}\ and\ \citenamefont
  {Dorning}(2003)}]{PhysRevE.68.026406}%
  \BibitemOpen
  \bibfield  {author} {\bibinfo {author} {\bibfnamefont {C.}~\bibnamefont
  {Lancellotti}}\ and\ \bibinfo {author} {\bibfnamefont {J.~J.}\ \bibnamefont
  {Dorning}},\ }\href {https://doi.org/10.1103/PhysRevE.68.026406} {\bibfield
  {journal} {\bibinfo  {journal} {Phys. Rev. E}\ }\textbf {\bibinfo {volume}
  {68}},\ \bibinfo {pages} {026406} (\bibinfo {year} {2003})}\BibitemShut
  {NoStop}%
\bibitem [{\citenamefont {Xu}\ \emph {et~al.}(2017)\citenamefont {Xu},
  \citenamefont {Sheng}, \citenamefont {Kong},\ and\ \citenamefont
  {Su}}]{xu2017effects}%
  \BibitemOpen
  \bibfield  {author} {\bibinfo {author} {\bibfnamefont {H.}~\bibnamefont
  {Xu}}, \bibinfo {author} {\bibfnamefont {Z.-M.}\ \bibnamefont {Sheng}},
  \bibinfo {author} {\bibfnamefont {X.-M.}\ \bibnamefont {Kong}},\ and\
  \bibinfo {author} {\bibfnamefont {F.-F.}\ \bibnamefont {Su}},\ }\href
  {https://doi.org/10.1063/1.4975020} {\bibfield  {journal} {\bibinfo
  {journal} {Physics of Plasmas}\ }\textbf {\bibinfo {volume} {24}},\ \bibinfo
  {pages} {022101} (\bibinfo {year} {2017})}\BibitemShut {NoStop}%
\bibitem [{\citenamefont {Tong}\ \emph {et~al.}(2018)\citenamefont {Tong},
  \citenamefont {Vasko}, \citenamefont {Mozer}, \citenamefont {Bale},
  \citenamefont {Roth}, \citenamefont {Artemyev}, \citenamefont {Ergun},
  \citenamefont {Giles}, \citenamefont {Lindqvist}, \citenamefont {Russell}
  \emph {et~al.}}]{tong2018simultaneous}%
  \BibitemOpen
  \bibfield  {author} {\bibinfo {author} {\bibfnamefont {Y.}~\bibnamefont
  {Tong}}, \bibinfo {author} {\bibfnamefont {I.}~\bibnamefont {Vasko}},
  \bibinfo {author} {\bibfnamefont {F.}~\bibnamefont {Mozer}}, \bibinfo
  {author} {\bibfnamefont {S.~D.}\ \bibnamefont {Bale}}, \bibinfo {author}
  {\bibfnamefont {I.}~\bibnamefont {Roth}}, \bibinfo {author} {\bibfnamefont
  {A.}~\bibnamefont {Artemyev}}, \bibinfo {author} {\bibfnamefont
  {R.}~\bibnamefont {Ergun}}, \bibinfo {author} {\bibfnamefont
  {B.}~\bibnamefont {Giles}}, \bibinfo {author} {\bibfnamefont {P.-A.}\
  \bibnamefont {Lindqvist}}, \bibinfo {author} {\bibfnamefont {C.}~\bibnamefont
  {Russell}}, \emph {et~al.},\ }\href {https://doi.org/10.1029/2018GL079044}
  {\bibfield  {journal} {\bibinfo  {journal} {Geophys. Res. Lett.}\ }\textbf
  {\bibinfo {volume} {45}},\ \bibinfo {pages} {11} (\bibinfo {year}
  {2018})}\BibitemShut {NoStop}%
\bibitem [{\citenamefont {Lotekar}\ \emph {et~al.}(2020)\citenamefont
  {Lotekar}, \citenamefont {Vasko}, \citenamefont {Mozer}, \citenamefont
  {Hutchinson}, \citenamefont {Artemyev}, \citenamefont {Bale}, \citenamefont
  {Bonnell}, \citenamefont {Ergun}, \citenamefont {Giles}, \citenamefont
  {Khotyaintsev} \emph {et~al.}}]{lotekar2020multisatellite}%
  \BibitemOpen
  \bibfield  {author} {\bibinfo {author} {\bibfnamefont {A.}~\bibnamefont
  {Lotekar}}, \bibinfo {author} {\bibfnamefont {I.}~\bibnamefont {Vasko}},
  \bibinfo {author} {\bibfnamefont {F.}~\bibnamefont {Mozer}}, \bibinfo
  {author} {\bibfnamefont {I.}~\bibnamefont {Hutchinson}}, \bibinfo {author}
  {\bibfnamefont {A.}~\bibnamefont {Artemyev}}, \bibinfo {author}
  {\bibfnamefont {S.}~\bibnamefont {Bale}}, \bibinfo {author} {\bibfnamefont
  {J.~W.}\ \bibnamefont {Bonnell}}, \bibinfo {author} {\bibfnamefont
  {R.}~\bibnamefont {Ergun}}, \bibinfo {author} {\bibfnamefont
  {B.}~\bibnamefont {Giles}}, \bibinfo {author} {\bibfnamefont {Y.~V.}\
  \bibnamefont {Khotyaintsev}}, \emph {et~al.},\ }\href
  {https://doi.org/10.1029/2020JA028066} {\bibfield  {journal} {\bibinfo
  {journal} {J. Geophys. Res. Space}\ }\textbf {\bibinfo {volume} {125}},\
  \bibinfo {pages} {e2020JA028066} (\bibinfo {year} {2020})}\BibitemShut
  {NoStop}%
\bibitem [{\citenamefont {Steinvall}\ \emph {et~al.}(2019)\citenamefont
  {Steinvall}, \citenamefont {Khotyaintsev}, \citenamefont {Graham},
  \citenamefont {Vaivads}, \citenamefont {Lindqvist}, \citenamefont {Russell},\
  and\ \citenamefont {Burch}}]{steinvall2019multispacecraft}%
  \BibitemOpen
  \bibfield  {author} {\bibinfo {author} {\bibfnamefont {K.}~\bibnamefont
  {Steinvall}}, \bibinfo {author} {\bibfnamefont {Y.~V.}\ \bibnamefont
  {Khotyaintsev}}, \bibinfo {author} {\bibfnamefont {D.~B.}\ \bibnamefont
  {Graham}}, \bibinfo {author} {\bibfnamefont {A.}~\bibnamefont {Vaivads}},
  \bibinfo {author} {\bibfnamefont {P.-A.}\ \bibnamefont {Lindqvist}}, \bibinfo
  {author} {\bibfnamefont {C.}~\bibnamefont {Russell}},\ and\ \bibinfo {author}
  {\bibfnamefont {J.}~\bibnamefont {Burch}},\ }\href
  {https://doi.org/10.1029/2018GL080757} {\bibfield  {journal} {\bibinfo
  {journal} {Geophys. Res. Lett.}\ }\textbf {\bibinfo {volume} {46}},\ \bibinfo
  {pages} {55} (\bibinfo {year} {2019})}\BibitemShut {NoStop}%
\bibitem [{\citenamefont {Mozer}\ \emph {et~al.}(2018)\citenamefont {Mozer},
  \citenamefont {Agapitov}, \citenamefont {Giles},\ and\ \citenamefont
  {Vasko}}]{mozer2018direct}%
  \BibitemOpen
  \bibfield  {author} {\bibinfo {author} {\bibfnamefont {F.}~\bibnamefont
  {Mozer}}, \bibinfo {author} {\bibfnamefont {O.}~\bibnamefont {Agapitov}},
  \bibinfo {author} {\bibfnamefont {B.}~\bibnamefont {Giles}},\ and\ \bibinfo
  {author} {\bibfnamefont {I.}~\bibnamefont {Vasko}},\ }\href
  {https://doi.org/10.1103/PhysRevLett.121.135102} {\bibfield  {journal}
  {\bibinfo  {journal} {Phys. Rev. Lett.}\ }\textbf {\bibinfo {volume} {121}},\
  \bibinfo {pages} {135102} (\bibinfo {year} {2018})}\BibitemShut {NoStop}%
\bibitem [{\citenamefont {Norgren}\ \emph {et~al.}(2015)\citenamefont
  {Norgren}, \citenamefont {Andr{\'e}}, \citenamefont {Vaivads},\ and\
  \citenamefont {Khotyaintsev}}]{norgren2015slow}%
  \BibitemOpen
  \bibfield  {author} {\bibinfo {author} {\bibfnamefont {C.}~\bibnamefont
  {Norgren}}, \bibinfo {author} {\bibfnamefont {M.}~\bibnamefont {Andr{\'e}}},
  \bibinfo {author} {\bibfnamefont {A.}~\bibnamefont {Vaivads}},\ and\ \bibinfo
  {author} {\bibfnamefont {Y.~V.}\ \bibnamefont {Khotyaintsev}},\ }\href
  {https://doi.org/10.1002/2015GL063218} {\bibfield  {journal} {\bibinfo
  {journal} {Geophys. Res. Lett.}\ }\textbf {\bibinfo {volume} {42}},\ \bibinfo
  {pages} {1654} (\bibinfo {year} {2015})}\BibitemShut {NoStop}%
\bibitem [{\citenamefont {Graham}\ \emph {et~al.}(2016)\citenamefont {Graham},
  \citenamefont {Khotyaintsev}, \citenamefont {Vaivads},\ and\ \citenamefont
  {Andre}}]{graham2016electrostatic}%
  \BibitemOpen
  \bibfield  {author} {\bibinfo {author} {\bibfnamefont {D.~B.}\ \bibnamefont
  {Graham}}, \bibinfo {author} {\bibfnamefont {Y.~V.}\ \bibnamefont
  {Khotyaintsev}}, \bibinfo {author} {\bibfnamefont {A.}~\bibnamefont
  {Vaivads}},\ and\ \bibinfo {author} {\bibfnamefont {M.}~\bibnamefont
  {Andre}},\ }\href {https://doi.org/10.1002/2015JA021527} {\bibfield
  {journal} {\bibinfo  {journal} {J. Geophys. Res. Space}\ }\textbf {\bibinfo
  {volume} {121}},\ \bibinfo {pages} {3069} (\bibinfo {year}
  {2016})}\BibitemShut {NoStop}%
\bibitem [{\citenamefont {Holmes}\ \emph {et~al.}(2018)\citenamefont {Holmes},
  \citenamefont {Ergun}, \citenamefont {Newman}, \citenamefont {Ahmadi},
  \citenamefont {Andersson}, \citenamefont {Le~Contel}, \citenamefont
  {Torbert}, \citenamefont {Giles}, \citenamefont {Strangeway},\ and\
  \citenamefont {Burch}}]{holmes2018electron}%
  \BibitemOpen
  \bibfield  {author} {\bibinfo {author} {\bibfnamefont {J.}~\bibnamefont
  {Holmes}}, \bibinfo {author} {\bibfnamefont {R.}~\bibnamefont {Ergun}},
  \bibinfo {author} {\bibfnamefont {D.}~\bibnamefont {Newman}}, \bibinfo
  {author} {\bibfnamefont {N.}~\bibnamefont {Ahmadi}}, \bibinfo {author}
  {\bibfnamefont {L.}~\bibnamefont {Andersson}}, \bibinfo {author}
  {\bibfnamefont {O.}~\bibnamefont {Le~Contel}}, \bibinfo {author}
  {\bibfnamefont {R.}~\bibnamefont {Torbert}}, \bibinfo {author} {\bibfnamefont
  {B.}~\bibnamefont {Giles}}, \bibinfo {author} {\bibfnamefont
  {R.}~\bibnamefont {Strangeway}},\ and\ \bibinfo {author} {\bibfnamefont
  {J.}~\bibnamefont {Burch}},\ }\href {https://doi.org/10.1029/2018JA025750}
  {\bibfield  {journal} {\bibinfo  {journal} {J. Geophys. Res. Space}\ }\textbf
  {\bibinfo {volume} {123}},\ \bibinfo {pages} {9963} (\bibinfo {year}
  {2018})}\BibitemShut {NoStop}%
\bibitem [{\citenamefont {Danielson}\ \emph {et~al.}(2004)\citenamefont
  {Danielson}, \citenamefont {Anderegg},\ and\ \citenamefont
  {Driscoll}}]{PhysRevLett.92.245003}%
  \BibitemOpen
  \bibfield  {author} {\bibinfo {author} {\bibfnamefont {J.~R.}\ \bibnamefont
  {Danielson}}, \bibinfo {author} {\bibfnamefont {F.}~\bibnamefont
  {Anderegg}},\ and\ \bibinfo {author} {\bibfnamefont {C.~F.}\ \bibnamefont
  {Driscoll}},\ }\href {https://doi.org/10.1103/PhysRevLett.92.245003}
  {\bibfield  {journal} {\bibinfo  {journal} {Phys. Rev. Lett.}\ }\textbf
  {\bibinfo {volume} {92}},\ \bibinfo {pages} {245003} (\bibinfo {year}
  {2004})}\BibitemShut {NoStop}%
\bibitem [{\citenamefont {Franklin}\ \emph {et~al.}(1972)\citenamefont
  {Franklin}, \citenamefont {Hamberger},\ and\ \citenamefont
  {Smith}}]{PhysRevLett.29.914}%
  \BibitemOpen
  \bibfield  {author} {\bibinfo {author} {\bibfnamefont {R.~N.}\ \bibnamefont
  {Franklin}}, \bibinfo {author} {\bibfnamefont {S.~M.}\ \bibnamefont
  {Hamberger}},\ and\ \bibinfo {author} {\bibfnamefont {G.~J.}\ \bibnamefont
  {Smith}},\ }\href {https://doi.org/10.1103/PhysRevLett.29.914} {\bibfield
  {journal} {\bibinfo  {journal} {Phys. Rev. Lett.}\ }\textbf {\bibinfo
  {volume} {29}},\ \bibinfo {pages} {914} (\bibinfo {year} {1972})}\BibitemShut
  {NoStop}%
\bibitem [{\citenamefont {Saeki}\ \emph {et~al.}(1979)\citenamefont {Saeki},
  \citenamefont {Michelsen}, \citenamefont {P{\'e}cseli},\ and\ \citenamefont
  {Rasmussen}}]{saeki1979formation}%
  \BibitemOpen
  \bibfield  {author} {\bibinfo {author} {\bibfnamefont {K.}~\bibnamefont
  {Saeki}}, \bibinfo {author} {\bibfnamefont {P.}~\bibnamefont {Michelsen}},
  \bibinfo {author} {\bibfnamefont {H.}~\bibnamefont {P{\'e}cseli}},\ and\
  \bibinfo {author} {\bibfnamefont {J.~J.}\ \bibnamefont {Rasmussen}},\ }\href
  {https://doi.org/10.1103/PhysRevLett.42.501} {\bibfield  {journal} {\bibinfo
  {journal} {Phys. Rev. Lett.}\ }\textbf {\bibinfo {volume} {42}},\ \bibinfo
  {pages} {501} (\bibinfo {year} {1979})}\BibitemShut {NoStop}%
\bibitem [{\citenamefont {Lefebvre}\ \emph {et~al.}(2010)\citenamefont
  {Lefebvre}, \citenamefont {Chen}, \citenamefont {Gekelman}, \citenamefont
  {Kintner}, \citenamefont {Pickett}, \citenamefont {Pribyl}, \citenamefont
  {Vincena}, \citenamefont {Chiang},\ and\ \citenamefont
  {Judy}}]{lefebvre2010laboratory}%
  \BibitemOpen
  \bibfield  {author} {\bibinfo {author} {\bibfnamefont {B.}~\bibnamefont
  {Lefebvre}}, \bibinfo {author} {\bibfnamefont {L.-J.}\ \bibnamefont {Chen}},
  \bibinfo {author} {\bibfnamefont {W.}~\bibnamefont {Gekelman}}, \bibinfo
  {author} {\bibfnamefont {P.}~\bibnamefont {Kintner}}, \bibinfo {author}
  {\bibfnamefont {J.}~\bibnamefont {Pickett}}, \bibinfo {author} {\bibfnamefont
  {P.}~\bibnamefont {Pribyl}}, \bibinfo {author} {\bibfnamefont
  {S.}~\bibnamefont {Vincena}}, \bibinfo {author} {\bibfnamefont
  {F.}~\bibnamefont {Chiang}},\ and\ \bibinfo {author} {\bibfnamefont
  {J.}~\bibnamefont {Judy}},\ }\href
  {https://doi.org/10.1103/PhysRevLett.105.115001} {\bibfield  {journal}
  {\bibinfo  {journal} {Phys. Rev. Lett.}\ }\textbf {\bibinfo {volume} {105}},\
  \bibinfo {pages} {115001} (\bibinfo {year} {2010})}\BibitemShut {NoStop}%
\bibitem [{\citenamefont {Rupp}\ \emph {et~al.}(2015)\citenamefont {Rupp},
  \citenamefont {L\'{o}pez},\ and\ \citenamefont {Araneda}}]{rupp2015density}%
  \BibitemOpen
  \bibfield  {author} {\bibinfo {author} {\bibfnamefont {C.~F.}\ \bibnamefont
  {Rupp}}, \bibinfo {author} {\bibfnamefont {R.~A.}\ \bibnamefont
  {L\'{o}pez}},\ and\ \bibinfo {author} {\bibfnamefont {J.~A.}\ \bibnamefont
  {Araneda}},\ }\href {https://doi.org/10.1063/1.4933022} {\bibfield  {journal}
  {\bibinfo  {journal} {Phys.~Plasmas}\ }\textbf {\bibinfo {volume} {22}},\
  \bibinfo {pages} {102306} (\bibinfo {year} {2015})}\BibitemShut {NoStop}%
\bibitem [{\citenamefont {Trivedi}\ and\ \citenamefont
  {Ganesh}(2016)}]{trivedi2016chirp}%
  \BibitemOpen
  \bibfield  {author} {\bibinfo {author} {\bibfnamefont {P.}~\bibnamefont
  {Trivedi}}\ and\ \bibinfo {author} {\bibfnamefont {R.}~\bibnamefont
  {Ganesh}},\ }\href {https://doi.org/10.1063/1.4953603} {\bibfield  {journal}
  {\bibinfo  {journal} {Phys. Plasmas}\ }\textbf {\bibinfo {volume} {23}},\
  \bibinfo {pages} {062112} (\bibinfo {year} {2016})}\BibitemShut {NoStop}%
\bibitem [{\citenamefont {Valentini}\ \emph {et~al.}(2005)\citenamefont
  {Valentini}, \citenamefont {Carbone}, \citenamefont {Veltri},\ and\
  \citenamefont {Mangeney}}]{PhysRevE.71.017402}%
  \BibitemOpen
  \bibfield  {author} {\bibinfo {author} {\bibfnamefont {F.}~\bibnamefont
  {Valentini}}, \bibinfo {author} {\bibfnamefont {V.}~\bibnamefont {Carbone}},
  \bibinfo {author} {\bibfnamefont {P.}~\bibnamefont {Veltri}},\ and\ \bibinfo
  {author} {\bibfnamefont {A.}~\bibnamefont {Mangeney}},\ }\href
  {https://doi.org/10.1103/PhysRevE.71.017402} {\bibfield  {journal} {\bibinfo
  {journal} {Phys. Rev. E}\ }\textbf {\bibinfo {volume} {71}},\ \bibinfo
  {pages} {017402} (\bibinfo {year} {2005})}\BibitemShut {NoStop}%
\bibitem [{\citenamefont {Valentini}(2005)}]{valentini2005nonlinear}%
  \BibitemOpen
  \bibfield  {author} {\bibinfo {author} {\bibfnamefont {F.}~\bibnamefont
  {Valentini}},\ }\href {https://doi.org/10.1063/1.1947967} {\bibfield
  {journal} {\bibinfo  {journal} {Phys. Plasmas}\ }\textbf {\bibinfo {volume}
  {12}},\ \bibinfo {pages} {072106} (\bibinfo {year} {2005})}\BibitemShut
  {NoStop}%
\bibitem [{\citenamefont {López}\ \emph {et~al.}(2017)\citenamefont {López},
  \citenamefont {Navarro}, \citenamefont {Pons},\ and\ \citenamefont
  {Araneda}}]{lopez2017landau}%
  \BibitemOpen
  \bibfield  {author} {\bibinfo {author} {\bibfnamefont {R.~A.}\ \bibnamefont
  {López}}, \bibinfo {author} {\bibfnamefont {R.~E.}\ \bibnamefont {Navarro}},
  \bibinfo {author} {\bibfnamefont {S.~I.}\ \bibnamefont {Pons}},\ and\
  \bibinfo {author} {\bibfnamefont {J.~A.}\ \bibnamefont {Araneda}},\ }\href
  {https://doi.org/10.1063/1.5004688} {\bibfield  {journal} {\bibinfo
  {journal} {Phys. Plasmas}\ }\textbf {\bibinfo {volume} {24}},\ \bibinfo
  {pages} {102119} (\bibinfo {year} {2017})}\BibitemShut {NoStop}%
\bibitem [{\citenamefont {Bernstein}\ \emph {et~al.}(1957)\citenamefont
  {Bernstein}, \citenamefont {Greene},\ and\ \citenamefont
  {Kruskal}}]{bernstein1957exact}%
  \BibitemOpen
  \bibfield  {author} {\bibinfo {author} {\bibfnamefont {I.~B.}\ \bibnamefont
  {Bernstein}}, \bibinfo {author} {\bibfnamefont {J.~M.}\ \bibnamefont
  {Greene}},\ and\ \bibinfo {author} {\bibfnamefont {M.~D.}\ \bibnamefont
  {Kruskal}},\ }\href {https://doi.org/10.1103/PhysRev.108.546} {\bibfield
  {journal} {\bibinfo  {journal} {Phys. Rev.}\ }\textbf {\bibinfo {volume}
  {108}},\ \bibinfo {pages} {546} (\bibinfo {year} {1957})}\BibitemShut
  {NoStop}%
\bibitem [{\citenamefont {Schamel}(1979)}]{schamel1979theory}%
  \BibitemOpen
  \bibfield  {author} {\bibinfo {author} {\bibfnamefont {H.}~\bibnamefont
  {Schamel}},\ }\href {https://doi.org/10.1088/0031-8949/20/3-4/006} {\bibfield
   {journal} {\bibinfo  {journal} {Phys. Scr.}\ }\textbf {\bibinfo {volume}
  {20}},\ \bibinfo {pages} {336} (\bibinfo {year} {1979})}\BibitemShut
  {NoStop}%
\bibitem [{\citenamefont {Schamel}(2012)}]{schamel2012cnoidal}%
  \BibitemOpen
  \bibfield  {author} {\bibinfo {author} {\bibfnamefont {H.}~\bibnamefont
  {Schamel}},\ }\href {https://doi.org/10.1063/1.3682047} {\bibfield  {journal}
  {\bibinfo  {journal} {Phys. Plasmas}\ }\textbf {\bibinfo {volume} {19}},\
  \bibinfo {pages} {020501} (\bibinfo {year} {2012})}\BibitemShut {NoStop}%
\bibitem [{\citenamefont {Schamel}(2015)}]{schamel2015particle}%
  \BibitemOpen
  \bibfield  {author} {\bibinfo {author} {\bibfnamefont {H.}~\bibnamefont
  {Schamel}},\ }\href {https://doi.org/10.1063/1.4916774} {\bibfield  {journal}
  {\bibinfo  {journal} {Phys. Plasmas}\ }\textbf {\bibinfo {volume} {22}},\
  \bibinfo {pages} {042301} (\bibinfo {year} {2015})}\BibitemShut {NoStop}%
\bibitem [{\citenamefont {Pankavich}\ and\ \citenamefont
  {Allen}(2014)}]{pankavich2014instability}%
  \BibitemOpen
  \bibfield  {author} {\bibinfo {author} {\bibfnamefont {S.}~\bibnamefont
  {Pankavich}}\ and\ \bibinfo {author} {\bibfnamefont {R.}~\bibnamefont
  {Allen}},\ }\href {https://doi.org/10.1140/epjd/e2014-50170-y} {\bibfield
  {journal} {\bibinfo  {journal} {Eur. Phys. J. D}\ }\textbf {\bibinfo {volume}
  {68}},\ \bibinfo {pages} {1} (\bibinfo {year} {2014})}\BibitemShut {NoStop}%
\bibitem [{\citenamefont {Manfredi}\ and\ \citenamefont
  {Bertrand}(2000)}]{manfredi2000stability}%
  \BibitemOpen
  \bibfield  {author} {\bibinfo {author} {\bibfnamefont {G.}~\bibnamefont
  {Manfredi}}\ and\ \bibinfo {author} {\bibfnamefont {P.}~\bibnamefont
  {Bertrand}},\ }\href {https://doi.org/10.1063/1.874081} {\bibfield  {journal}
  {\bibinfo  {journal} {Phys. Plasmas}\ }\textbf {\bibinfo {volume} {7}},\
  \bibinfo {pages} {2425} (\bibinfo {year} {2000})}\BibitemShut {NoStop}%
\bibitem [{\citenamefont {Morse}\ and\ \citenamefont
  {Nielson}(1969)}]{morse1969one}%
  \BibitemOpen
  \bibfield  {author} {\bibinfo {author} {\bibfnamefont {R.~L.}\ \bibnamefont
  {Morse}}\ and\ \bibinfo {author} {\bibfnamefont {C.}~\bibnamefont
  {Nielson}},\ }\href {https://doi.org/10.1103/PhysRevLett.23.1087} {\bibfield
  {journal} {\bibinfo  {journal} {Phys. Rev. Lett.}\ }\textbf {\bibinfo
  {volume} {23}},\ \bibinfo {pages} {1087} (\bibinfo {year}
  {1969})}\BibitemShut {NoStop}%
\bibitem [{\citenamefont {Galeotti}\ and\ \citenamefont
  {Califano}(2005)}]{galeotti2005asymptotic}%
  \BibitemOpen
  \bibfield  {author} {\bibinfo {author} {\bibfnamefont {L.}~\bibnamefont
  {Galeotti}}\ and\ \bibinfo {author} {\bibfnamefont {F.}~\bibnamefont
  {Califano}},\ }\href {https://doi.org/10.1103/PhysRevLett.95.015002}
  {\bibfield  {journal} {\bibinfo  {journal} {Phys. Rev. Lett.}\ }\textbf
  {\bibinfo {volume} {95}},\ \bibinfo {pages} {015002} (\bibinfo {year}
  {2005})}\BibitemShut {NoStop}%
\bibitem [{\citenamefont {Califano}\ \emph {et~al.}(2006)\citenamefont
  {Califano}, \citenamefont {Galeotti},\ and\ \citenamefont
  {Mangeney}}]{califano2006vlasov}%
  \BibitemOpen
  \bibfield  {author} {\bibinfo {author} {\bibfnamefont {F.}~\bibnamefont
  {Califano}}, \bibinfo {author} {\bibfnamefont {L.}~\bibnamefont {Galeotti}},\
  and\ \bibinfo {author} {\bibfnamefont {A.}~\bibnamefont {Mangeney}},\ }\href
  {https://doi.org/10.1063/1.2215596} {\bibfield  {journal} {\bibinfo
  {journal} {Phys. Plasmas}\ }\textbf {\bibinfo {volume} {13}},\ \bibinfo
  {pages} {082102} (\bibinfo {year} {2006})}\BibitemShut {NoStop}%
\bibitem [{\citenamefont {Shoucri}(2017)}]{shoucri2017formation}%
  \BibitemOpen
  \bibfield  {author} {\bibinfo {author} {\bibfnamefont {M.}~\bibnamefont
  {Shoucri}},\ }\href {https://doi.org/10.1017/S0263034617000775} {\bibfield
  {journal} {\bibinfo  {journal} {Laser Part. Beams}\ }\textbf {\bibinfo
  {volume} {35}},\ \bibinfo {pages} {706} (\bibinfo {year} {2017})}\BibitemShut
  {NoStop}%
\bibitem [{\citenamefont {Brunetti}\ \emph {et~al.}(2000)\citenamefont
  {Brunetti}, \citenamefont {Califano},\ and\ \citenamefont
  {Pegoraro}}]{brunetti2000asymptotic}%
  \BibitemOpen
  \bibfield  {author} {\bibinfo {author} {\bibfnamefont {M.}~\bibnamefont
  {Brunetti}}, \bibinfo {author} {\bibfnamefont {F.}~\bibnamefont {Califano}},\
  and\ \bibinfo {author} {\bibfnamefont {F.}~\bibnamefont {Pegoraro}},\ }\href
  {https://doi.org/10.1103/PhysRevE.62.4109} {\bibfield  {journal} {\bibinfo
  {journal} {Phys. Rev. E}\ }\textbf {\bibinfo {volume} {62}},\ \bibinfo
  {pages} {4109} (\bibinfo {year} {2000})}\BibitemShut {NoStop}%
\bibitem [{\citenamefont {Pandey}\ and\ \citenamefont
  {Ganesh}(2021)}]{pandey2021trapped}%
  \BibitemOpen
  \bibfield  {author} {\bibinfo {author} {\bibfnamefont {S.~K.}\ \bibnamefont
  {Pandey}}\ and\ \bibinfo {author} {\bibfnamefont {R.}~\bibnamefont
  {Ganesh}},\ }\href {https://doi.org/10.1088/1402-4896/ac25a1} {\bibfield
  {journal} {\bibinfo  {journal} {Phys. Scr.}\ }\textbf {\bibinfo {volume}
  {96}},\ \bibinfo {pages} {125616} (\bibinfo {year} {2021})}\BibitemShut
  {NoStop}%
\bibitem [{\citenamefont {Ghizzo}\ \emph {et~al.}(1988)\citenamefont {Ghizzo},
  \citenamefont {Izrar}, \citenamefont {Bertrand}, \citenamefont {Fijalkow},
  \citenamefont {Feix},\ and\ \citenamefont {Shoucri}}]{ghizzo1988stability}%
  \BibitemOpen
  \bibfield  {author} {\bibinfo {author} {\bibfnamefont {A.}~\bibnamefont
  {Ghizzo}}, \bibinfo {author} {\bibfnamefont {B.}~\bibnamefont {Izrar}},
  \bibinfo {author} {\bibfnamefont {P.}~\bibnamefont {Bertrand}}, \bibinfo
  {author} {\bibfnamefont {E.}~\bibnamefont {Fijalkow}}, \bibinfo {author}
  {\bibfnamefont {M.}~\bibnamefont {Feix}},\ and\ \bibinfo {author}
  {\bibfnamefont {M.}~\bibnamefont {Shoucri}},\ }\href
  {https://doi.org/10.1063/1.866579} {\bibfield  {journal} {\bibinfo  {journal}
  {Phys. Fluids}\ }\textbf {\bibinfo {volume} {31}},\ \bibinfo {pages} {72}
  (\bibinfo {year} {1988})}\BibitemShut {NoStop}%
\bibitem [{\citenamefont {Buchanan}\ and\ \citenamefont
  {Dorning}(1993)}]{PhysRevLett.70.3732}%
  \BibitemOpen
  \bibfield  {author} {\bibinfo {author} {\bibfnamefont {M.}~\bibnamefont
  {Buchanan}}\ and\ \bibinfo {author} {\bibfnamefont {J.~J.}\ \bibnamefont
  {Dorning}},\ }\href {https://doi.org/10.1103/PhysRevLett.70.3732} {\bibfield
  {journal} {\bibinfo  {journal} {Phys. Rev. Lett.}\ }\textbf {\bibinfo
  {volume} {70}},\ \bibinfo {pages} {3732} (\bibinfo {year}
  {1993})}\BibitemShut {NoStop}%
\bibitem [{\citenamefont {Buchanan}\ and\ \citenamefont
  {Dorning}(1994)}]{PhysRevE.50.1465}%
  \BibitemOpen
  \bibfield  {author} {\bibinfo {author} {\bibfnamefont {M.}~\bibnamefont
  {Buchanan}}\ and\ \bibinfo {author} {\bibfnamefont {J.}~\bibnamefont
  {Dorning}},\ }\href {https://doi.org/10.1103/PhysRevE.50.1465} {\bibfield
  {journal} {\bibinfo  {journal} {Phys. Rev. E}\ }\textbf {\bibinfo {volume}
  {50}},\ \bibinfo {pages} {1465} (\bibinfo {year} {1994})}\BibitemShut
  {NoStop}%
\bibitem [{\citenamefont {Demeio}\ and\ \citenamefont
  {Zweifel}(1990)}]{demeio1990numerical}%
  \BibitemOpen
  \bibfield  {author} {\bibinfo {author} {\bibfnamefont {L.}~\bibnamefont
  {Demeio}}\ and\ \bibinfo {author} {\bibfnamefont {P.}~\bibnamefont
  {Zweifel}},\ }\href {https://doi.org/10.1063/1.859265} {\bibfield  {journal}
  {\bibinfo  {journal} {Phys. Plasmas}\ }\textbf {\bibinfo {volume} {2}},\
  \bibinfo {pages} {1252} (\bibinfo {year} {1990})}\BibitemShut {NoStop}%
\bibitem [{\citenamefont {Demeio}(2001)}]{demeio2001numerical}%
  \BibitemOpen
  \bibfield  {author} {\bibinfo {author} {\bibfnamefont {L.}~\bibnamefont
  {Demeio}},\ }\href {https://doi.org/10.1081/TT-100105932} {\bibfield
  {journal} {\bibinfo  {journal} {Transp. Theory Stat. Phys.}\ }\textbf
  {\bibinfo {volume} {30}},\ \bibinfo {pages} {457} (\bibinfo {year}
  {2001})}\BibitemShut {NoStop}%
\bibitem [{\citenamefont {Fijalkow}\ and\ \citenamefont
  {Nocera}(2005)}]{fijalkow2005superposition}%
  \BibitemOpen
  \bibfield  {author} {\bibinfo {author} {\bibfnamefont {E.}~\bibnamefont
  {Fijalkow}}\ and\ \bibinfo {author} {\bibfnamefont {L.}~\bibnamefont
  {Nocera}},\ }\href {https://doi.org/10.1017/S0022377804003368} {\bibfield
  {journal} {\bibinfo  {journal} {J. Plasma Phys.}\ }\textbf {\bibinfo {volume}
  {71}},\ \bibinfo {pages} {401} (\bibinfo {year} {2005})}\BibitemShut
  {NoStop}%
\bibitem [{\citenamefont {Lancellotti}\ and\ \citenamefont
  {Dorning}(1999)}]{lancellotti1999time}%
  \BibitemOpen
  \bibfield  {author} {\bibinfo {author} {\bibfnamefont {C.}~\bibnamefont
  {Lancellotti}}\ and\ \bibinfo {author} {\bibfnamefont {J.}~\bibnamefont
  {Dorning}},\ }\href {https://doi.org/10.1063/1.532932} {\bibfield  {journal}
  {\bibinfo  {journal} {J. Math. Phys.}\ }\textbf {\bibinfo {volume} {40}},\
  \bibinfo {pages} {3895} (\bibinfo {year} {1999})}\BibitemShut {NoStop}%
\bibitem [{\citenamefont {Lancellotti}\ and\ \citenamefont
  {Dorning}(2009)}]{lancellotti2009nonlinear}%
  \BibitemOpen
  \bibfield  {author} {\bibinfo {author} {\bibfnamefont {C.}~\bibnamefont
  {Lancellotti}}\ and\ \bibinfo {author} {\bibfnamefont {J.}~\bibnamefont
  {Dorning}},\ }\href {https://doi.org/10.1080/00411450903081313} {\bibfield
  {journal} {\bibinfo  {journal} {Transp. Theory Stat. Phys.}\ }\textbf
  {\bibinfo {volume} {38}},\ \bibinfo {pages} {1} (\bibinfo {year}
  {2009})}\BibitemShut {NoStop}%
\bibitem [{\citenamefont {Bezanson}\ \emph {et~al.}(2017)\citenamefont
  {Bezanson}, \citenamefont {Edelman}, \citenamefont {Karpinski},\ and\
  \citenamefont {Shah}}]{Julia}%
  \BibitemOpen
  \bibfield  {author} {\bibinfo {author} {\bibfnamefont {J.}~\bibnamefont
  {Bezanson}}, \bibinfo {author} {\bibfnamefont {A.}~\bibnamefont {Edelman}},
  \bibinfo {author} {\bibfnamefont {S.}~\bibnamefont {Karpinski}},\ and\
  \bibinfo {author} {\bibfnamefont {V.~B.}\ \bibnamefont {Shah}},\ }\href
  {https://doi.org/10.1137/141000671} {\bibfield  {journal} {\bibinfo
  {journal} {SIAM review}\ }\textbf {\bibinfo {volume} {59}},\ \bibinfo {pages}
  {65} (\bibinfo {year} {2017})}\BibitemShut {NoStop}%
\bibitem [{\citenamefont {Sonnendr{\"u}cker}\ \emph {et~al.}(1999)\citenamefont
  {Sonnendr{\"u}cker}, \citenamefont {Roche}, \citenamefont {Bertrand},\ and\
  \citenamefont {Ghizzo}}]{sonnendrucker1999semi}%
  \BibitemOpen
  \bibfield  {author} {\bibinfo {author} {\bibfnamefont {E.}~\bibnamefont
  {Sonnendr{\"u}cker}}, \bibinfo {author} {\bibfnamefont {J.}~\bibnamefont
  {Roche}}, \bibinfo {author} {\bibfnamefont {P.}~\bibnamefont {Bertrand}},\
  and\ \bibinfo {author} {\bibfnamefont {A.}~\bibnamefont {Ghizzo}},\ }\href
  {https://doi.org/10.1006/jcph.1998.6148} {\bibfield  {journal} {\bibinfo
  {journal} {J. Comput. Phys.}\ }\textbf {\bibinfo {volume} {149}},\ \bibinfo
  {pages} {201} (\bibinfo {year} {1999})}\BibitemShut {NoStop}%
\bibitem [{\citenamefont {McLachlan}\ and\ \citenamefont
  {Atela}(1992)}]{mclachlan1992}%
  \BibitemOpen
  \bibfield  {author} {\bibinfo {author} {\bibfnamefont {R.~I.}\ \bibnamefont
  {McLachlan}}\ and\ \bibinfo {author} {\bibfnamefont {P.}~\bibnamefont
  {Atela}},\ }\href {https://doi.org/10.1088/0951-7715/5/2/011} {\bibfield
  {journal} {\bibinfo  {journal} {Nonlinearity}\ }\textbf {\bibinfo {volume}
  {5}},\ \bibinfo {pages} {541} (\bibinfo {year} {1992})}\BibitemShut {NoStop}%
\bibitem [{\citenamefont {Fornberg}\ and\ \citenamefont
  {Sloan}(1994)}]{fornberg1994review}%
  \BibitemOpen
  \bibfield  {author} {\bibinfo {author} {\bibfnamefont {B.}~\bibnamefont
  {Fornberg}}\ and\ \bibinfo {author} {\bibfnamefont {D.~M.}\ \bibnamefont
  {Sloan}},\ }\href {https://doi.org/10.1017/S0962492900002440} {\bibfield
  {journal} {\bibinfo  {journal} {Acta Numer.}\ }\textbf {\bibinfo {volume}
  {3}},\ \bibinfo {pages} {203} (\bibinfo {year} {1994})}\BibitemShut {NoStop}%
\bibitem [{\citenamefont {Boyd}(2001)}]{boyd2001chebyshev}%
  \BibitemOpen
  \bibfield  {author} {\bibinfo {author} {\bibfnamefont {J.~P.}\ \bibnamefont
  {Boyd}},\ }\href@noop {} {\emph {\bibinfo {title} {Chebyshev and Fourier
  spectral methods}}}\ (\bibinfo {year} {2001})\BibitemShut {NoStop}%
\bibitem [{\citenamefont {Parker}\ and\ \citenamefont
  {Dellar}(2015)}]{parker2015fourier}%
  \BibitemOpen
  \bibfield  {author} {\bibinfo {author} {\bibfnamefont {J.~T.}\ \bibnamefont
  {Parker}}\ and\ \bibinfo {author} {\bibfnamefont {P.~J.}\ \bibnamefont
  {Dellar}},\ }\bibfield  {journal} {\bibinfo  {journal} {J. Plasma Phys.}\
  }\textbf {\bibinfo {volume} {81}},\ \href
  {https://doi.org/10.1017/S0022377814001287} {10.1017/S0022377814001287}
  (\bibinfo {year} {2015})\BibitemShut {NoStop}%
\bibitem [{\citenamefont {O'Neil}\ \emph {et~al.}(1971)\citenamefont {O'Neil},
  \citenamefont {Winfrey},\ and\ \citenamefont {Malmberg}}]{o1971nonlinear}%
  \BibitemOpen
  \bibfield  {author} {\bibinfo {author} {\bibfnamefont {T.~M.}\ \bibnamefont
  {O'Neil}}, \bibinfo {author} {\bibfnamefont {J.}~\bibnamefont {Winfrey}},\
  and\ \bibinfo {author} {\bibfnamefont {J.}~\bibnamefont {Malmberg}},\ }\href
  {https://doi.org/10.1063/1.1693587} {\bibfield  {journal} {\bibinfo
  {journal} {Phys. Fluids}\ }\textbf {\bibinfo {volume} {14}},\ \bibinfo
  {pages} {1204} (\bibinfo {year} {1971})}\BibitemShut {NoStop}%
\bibitem [{\citenamefont {Hara}\ \emph {et~al.}(2015)\citenamefont {Hara},
  \citenamefont {Chapman}, \citenamefont {Banks}, \citenamefont {Brunner},
  \citenamefont {Joseph}, \citenamefont {Berger},\ and\ \citenamefont
  {Boyd}}]{Hara2015bunching}%
  \BibitemOpen
  \bibfield  {author} {\bibinfo {author} {\bibfnamefont {K.}~\bibnamefont
  {Hara}}, \bibinfo {author} {\bibfnamefont {T.}~\bibnamefont {Chapman}},
  \bibinfo {author} {\bibfnamefont {J.~W.}\ \bibnamefont {Banks}}, \bibinfo
  {author} {\bibfnamefont {S.}~\bibnamefont {Brunner}}, \bibinfo {author}
  {\bibfnamefont {I.}~\bibnamefont {Joseph}}, \bibinfo {author} {\bibfnamefont
  {R.~L.}\ \bibnamefont {Berger}},\ and\ \bibinfo {author} {\bibfnamefont
  {I.~D.}\ \bibnamefont {Boyd}},\ }\href {https://doi.org/10.1063/1.4906884}
  {\bibfield  {journal} {\bibinfo  {journal} {Phys.~Plasmas}\ }\textbf
  {\bibinfo {volume} {22}},\ \bibinfo {pages} {022104} (\bibinfo {year}
  {2015})}\BibitemShut {NoStop}%
\bibitem [{\citenamefont {Brunner}\ \emph {et~al.}(2014)\citenamefont
  {Brunner}, \citenamefont {Berger}, \citenamefont {Cohen}, \citenamefont
  {Hausammann},\ and\ \citenamefont {Valeo}}]{brunner2014kinetic}%
  \BibitemOpen
  \bibfield  {author} {\bibinfo {author} {\bibfnamefont {S.}~\bibnamefont
  {Brunner}}, \bibinfo {author} {\bibfnamefont {R.}~\bibnamefont {Berger}},
  \bibinfo {author} {\bibfnamefont {B.}~\bibnamefont {Cohen}}, \bibinfo
  {author} {\bibfnamefont {L.}~\bibnamefont {Hausammann}},\ and\ \bibinfo
  {author} {\bibfnamefont {E.}~\bibnamefont {Valeo}},\ }\href
  {https://doi.org/10.1063/1.4896753} {\bibfield  {journal} {\bibinfo
  {journal} {Phys. Plasmas}\ }\textbf {\bibinfo {volume} {21}},\ \bibinfo
  {pages} {102104} (\bibinfo {year} {2014})}\BibitemShut {NoStop}%
\bibitem [{\citenamefont {Dodin}\ \emph {et~al.}(2013)\citenamefont {Dodin},
  \citenamefont {Schmit}, \citenamefont {Rocks},\ and\ \citenamefont
  {Fisch}}]{Dodin2013negativemass}%
  \BibitemOpen
  \bibfield  {author} {\bibinfo {author} {\bibfnamefont {I.~Y.}\ \bibnamefont
  {Dodin}}, \bibinfo {author} {\bibfnamefont {P.~F.}\ \bibnamefont {Schmit}},
  \bibinfo {author} {\bibfnamefont {J.}~\bibnamefont {Rocks}},\ and\ \bibinfo
  {author} {\bibfnamefont {N.~J.}\ \bibnamefont {Fisch}},\ }\href
  {https://doi.org/10.1103/PhysRevLett.110.215006} {\bibfield  {journal}
  {\bibinfo  {journal} {Phys. Rev. Lett.}\ }\textbf {\bibinfo {volume} {110}},\
  \bibinfo {pages} {215006} (\bibinfo {year} {2013})}\BibitemShut {NoStop}%
\bibitem [{\citenamefont {O'neil}\ and\ \citenamefont
  {Winfrey}(1972)}]{o1972nonlinear}%
  \BibitemOpen
  \bibfield  {author} {\bibinfo {author} {\bibfnamefont {T.}~\bibnamefont
  {O'neil}}\ and\ \bibinfo {author} {\bibfnamefont {J.}~\bibnamefont
  {Winfrey}},\ }\href {https://doi.org/10.1063/1.1694117} {\bibfield  {journal}
  {\bibinfo  {journal} {Phys. Fluids}\ }\textbf {\bibinfo {volume} {15}},\
  \bibinfo {pages} {1514} (\bibinfo {year} {1972})}\BibitemShut {NoStop}%
\bibitem [{\citenamefont {Bouchoule}\ and\ \citenamefont
  {Weinfeld}(1976)}]{bouchoule1976generation}%
  \BibitemOpen
  \bibfield  {author} {\bibinfo {author} {\bibfnamefont {A.}~\bibnamefont
  {Bouchoule}}\ and\ \bibinfo {author} {\bibfnamefont {M.}~\bibnamefont
  {Weinfeld}},\ }\href {https://doi.org/10.1103/PhysRevLett.36.1144} {\bibfield
   {journal} {\bibinfo  {journal} {Phys. Rev. Lett.}\ }\textbf {\bibinfo
  {volume} {36}},\ \bibinfo {pages} {1144} (\bibinfo {year}
  {1976})}\BibitemShut {NoStop}%
\bibitem [{\citenamefont {Firpo}\ \emph {et~al.}(2001)\citenamefont {Firpo},
  \citenamefont {Doveil}, \citenamefont {Elskens}, \citenamefont {Bertrand},
  \citenamefont {Poleni},\ and\ \citenamefont {Guyomarc’h}}]{firpo2001long}%
  \BibitemOpen
  \bibfield  {author} {\bibinfo {author} {\bibfnamefont {M.-C.}\ \bibnamefont
  {Firpo}}, \bibinfo {author} {\bibfnamefont {F.}~\bibnamefont {Doveil}},
  \bibinfo {author} {\bibfnamefont {Y.}~\bibnamefont {Elskens}}, \bibinfo
  {author} {\bibfnamefont {P.}~\bibnamefont {Bertrand}}, \bibinfo {author}
  {\bibfnamefont {M.}~\bibnamefont {Poleni}},\ and\ \bibinfo {author}
  {\bibfnamefont {D.}~\bibnamefont {Guyomarc’h}},\ }\href
  {https://doi.org/10.1103/PhysRevE.64.026407} {\bibfield  {journal} {\bibinfo
  {journal} {Phys. Rev. E}\ }\textbf {\bibinfo {volume} {64}},\ \bibinfo
  {pages} {026407} (\bibinfo {year} {2001})}\BibitemShut {NoStop}%
\bibitem [{\citenamefont {Carlevaro}\ \emph {et~al.}(2015)\citenamefont
  {Carlevaro}, \citenamefont {Falessi}, \citenamefont {Montani},\ and\
  \citenamefont {Zonca}}]{carlevaro2015nonlinear}%
  \BibitemOpen
  \bibfield  {author} {\bibinfo {author} {\bibfnamefont {N.}~\bibnamefont
  {Carlevaro}}, \bibinfo {author} {\bibfnamefont {M.~V.}\ \bibnamefont
  {Falessi}}, \bibinfo {author} {\bibfnamefont {G.}~\bibnamefont {Montani}},\
  and\ \bibinfo {author} {\bibfnamefont {F.}~\bibnamefont {Zonca}},\ }\bibfield
   {journal} {\bibinfo  {journal} {J. Plasma Phys.}\ }\textbf {\bibinfo
  {volume} {81}},\ \href {https://doi.org/10.1017/S0022377815001002}
  {10.1017/S0022377815001002} (\bibinfo {year} {2015})\BibitemShut {NoStop}%
\bibitem [{\citenamefont {Tennyson}\ \emph {et~al.}(1994)\citenamefont
  {Tennyson}, \citenamefont {Meiss},\ and\ \citenamefont
  {Morrison}}]{tennyson1994self}%
  \BibitemOpen
  \bibfield  {author} {\bibinfo {author} {\bibfnamefont {J.}~\bibnamefont
  {Tennyson}}, \bibinfo {author} {\bibfnamefont {J.}~\bibnamefont {Meiss}},\
  and\ \bibinfo {author} {\bibfnamefont {P.}~\bibnamefont {Morrison}},\ }\href
  {https://doi.org/10.1016/0167-2789(94)90178-3} {\bibfield  {journal}
  {\bibinfo  {journal} {Physica D}\ }\textbf {\bibinfo {volume} {71}},\
  \bibinfo {pages} {1} (\bibinfo {year} {1994})}\BibitemShut {NoStop}%
\bibitem [{\citenamefont {Farina}\ and\ \citenamefont
  {Pozzoli}(2004)}]{farina2004large}%
  \BibitemOpen
  \bibfield  {author} {\bibinfo {author} {\bibfnamefont {D.}~\bibnamefont
  {Farina}}\ and\ \bibinfo {author} {\bibfnamefont {R.}~\bibnamefont
  {Pozzoli}},\ }\href {https://doi.org/10.1103/PhysRevE.70.036407} {\bibfield
  {journal} {\bibinfo  {journal} {Phys. Rev. E}\ }\textbf {\bibinfo {volume}
  {70}},\ \bibinfo {pages} {036407} (\bibinfo {year} {2004})}\BibitemShut
  {NoStop}%
\bibitem [{\citenamefont {Adam}\ \emph {et~al.}(1981)\citenamefont {Adam},
  \citenamefont {Laval},\ and\ \citenamefont {Mendonca}}]{adam1981time}%
  \BibitemOpen
  \bibfield  {author} {\bibinfo {author} {\bibfnamefont {J.}~\bibnamefont
  {Adam}}, \bibinfo {author} {\bibfnamefont {G.}~\bibnamefont {Laval}},\ and\
  \bibinfo {author} {\bibfnamefont {I.}~\bibnamefont {Mendonca}},\ }\href
  {https://doi.org/10.1063/1.863354} {\bibfield  {journal} {\bibinfo  {journal}
  {Phys. Fluids}\ }\textbf {\bibinfo {volume} {24}},\ \bibinfo {pages} {260}
  (\bibinfo {year} {1981})}\BibitemShut {NoStop}%
\bibitem [{\citenamefont {Hou}\ \emph {et~al.}(2015)\citenamefont {Hou},
  \citenamefont {Chen}, \citenamefont {Yu}, \citenamefont {Wu},\ and\
  \citenamefont {Wu}}]{hou2015excitation}%
  \BibitemOpen
  \bibfield  {author} {\bibinfo {author} {\bibfnamefont {Y.}~\bibnamefont
  {Hou}}, \bibinfo {author} {\bibfnamefont {M.}~\bibnamefont {Chen}}, \bibinfo
  {author} {\bibfnamefont {M.}~\bibnamefont {Yu}}, \bibinfo {author}
  {\bibfnamefont {B.}~\bibnamefont {Wu}},\ and\ \bibinfo {author}
  {\bibfnamefont {Y.}~\bibnamefont {Wu}},\ }\href
  {https://doi.org/10.1063/1.4936267} {\bibfield  {journal} {\bibinfo
  {journal} {Phys. Plasmas}\ }\textbf {\bibinfo {volume} {22}},\ \bibinfo
  {pages} {122101} (\bibinfo {year} {2015})}\BibitemShut {NoStop}%
\bibitem [{\citenamefont {Dupree}(1982)}]{dupree1982theory}%
  \BibitemOpen
  \bibfield  {author} {\bibinfo {author} {\bibfnamefont {T.~H.}\ \bibnamefont
  {Dupree}},\ }\href {https://doi.org/10.1063/1.863734} {\bibfield  {journal}
  {\bibinfo  {journal} {Phys. Fluids}\ }\textbf {\bibinfo {volume} {25}},\
  \bibinfo {pages} {277} (\bibinfo {year} {1982})}\BibitemShut {NoStop}%
\bibitem [{\citenamefont {Dupree}(1983)}]{dupree1983growth}%
  \BibitemOpen
  \bibfield  {author} {\bibinfo {author} {\bibfnamefont {T.~H.}\ \bibnamefont
  {Dupree}},\ }\href {https://doi.org/10.1063/1.864430} {\bibfield  {journal}
  {\bibinfo  {journal} {Phys. Fluids}\ }\textbf {\bibinfo {volume} {26}},\
  \bibinfo {pages} {2460} (\bibinfo {year} {1983})}\BibitemShut {NoStop}%
\bibitem [{\citenamefont {Berman}\ \emph {et~al.}(1985)\citenamefont {Berman},
  \citenamefont {Tetreault},\ and\ \citenamefont
  {Dupree}}]{berman1985simulation}%
  \BibitemOpen
  \bibfield  {author} {\bibinfo {author} {\bibfnamefont {R.~H.}\ \bibnamefont
  {Berman}}, \bibinfo {author} {\bibfnamefont {D.~J.}\ \bibnamefont
  {Tetreault}},\ and\ \bibinfo {author} {\bibfnamefont {T.~H.}\ \bibnamefont
  {Dupree}},\ }\href {https://doi.org/10.1063/1.865176} {\bibfield  {journal}
  {\bibinfo  {journal} {Phys. Fluids}\ }\textbf {\bibinfo {volume} {28}},\
  \bibinfo {pages} {155} (\bibinfo {year} {1985})}\BibitemShut {NoStop}%
\bibitem [{\citenamefont {Schamel}\ \emph {et~al.}(2017)\citenamefont
  {Schamel}, \citenamefont {Mandal},\ and\ \citenamefont
  {Sharma}}]{schamel2017nonlinear}%
  \BibitemOpen
  \bibfield  {author} {\bibinfo {author} {\bibfnamefont {H.}~\bibnamefont
  {Schamel}}, \bibinfo {author} {\bibfnamefont {D.}~\bibnamefont {Mandal}},\
  and\ \bibinfo {author} {\bibfnamefont {D.}~\bibnamefont {Sharma}},\ }\href
  {https://doi.org/10.1063/1.4978477} {\bibfield  {journal} {\bibinfo
  {journal} {Physics of Plasmas}\ }\textbf {\bibinfo {volume} {24}},\ \bibinfo
  {pages} {032109} (\bibinfo {year} {2017})}\BibitemShut {NoStop}%
\bibitem [{\citenamefont {Yoon}(2007)}]{Yoon2007}%
  \BibitemOpen
  \bibfield  {author} {\bibinfo {author} {\bibfnamefont {P.~H.}\ \bibnamefont
  {Yoon}},\ }\href {https://doi.org/10.1063/1.2741388} {\bibfield  {journal}
  {\bibinfo  {journal} {Phys.~Plasmas}\ }\textbf {\bibinfo {volume} {14}},\
  \bibinfo {pages} {064504} (\bibinfo {year} {2007})}\BibitemShut {NoStop}%
\bibitem [{\citenamefont {Klimas}\ and\ \citenamefont
  {Vi{\~n}as}(2018)}]{klimas2018absence}%
  \BibitemOpen
  \bibfield  {author} {\bibinfo {author} {\bibfnamefont {A.~J.}\ \bibnamefont
  {Klimas}}\ and\ \bibinfo {author} {\bibfnamefont {A.~F.}\ \bibnamefont
  {Vi{\~n}as}},\ }\bibfield  {journal} {\bibinfo  {journal} {J. Plasma Phys.}\
  }\textbf {\bibinfo {volume} {84}},\ \href
  {https://doi.org/10.1017/S0022377818000776} {10.1017/S0022377818000776}
  (\bibinfo {year} {2018})\BibitemShut {NoStop}%
\bibitem [{\citenamefont {Vi{\~n}as}\ and\ \citenamefont
  {Klimas}(2018)}]{vinas2018flux}%
  \BibitemOpen
  \bibfield  {author} {\bibinfo {author} {\bibfnamefont {A.~F.}\ \bibnamefont
  {Vi{\~n}as}}\ and\ \bibinfo {author} {\bibfnamefont {A.~J.}\ \bibnamefont
  {Klimas}},\ }\href {https://doi.org/10.1016/j.jcp.2018.09.014} {\bibfield
  {journal} {\bibinfo  {journal} {J. Comput. Phys.}\ }\textbf {\bibinfo
  {volume} {375}},\ \bibinfo {pages} {983} (\bibinfo {year}
  {2018})}\BibitemShut {NoStop}%
\bibitem [{\citenamefont {Besse}\ and\ \citenamefont
  {Sonnendr{\"u}cker}(2003)}]{besse2003semi}%
  \BibitemOpen
  \bibfield  {author} {\bibinfo {author} {\bibfnamefont {N.}~\bibnamefont
  {Besse}}\ and\ \bibinfo {author} {\bibfnamefont {E.}~\bibnamefont
  {Sonnendr{\"u}cker}},\ }\href {https://doi.org/10.1016/S0021-9991(03)00318-8}
  {\bibfield  {journal} {\bibinfo  {journal} {J. Comput. Phys.}\ }\textbf
  {\bibinfo {volume} {191}},\ \bibinfo {pages} {341} (\bibinfo {year}
  {2003})}\BibitemShut {NoStop}%
\bibitem [{\citenamefont {Einkemmer}\ and\ \citenamefont
  {Lubich}(2018)}]{einkemmer2018low}%
  \BibitemOpen
  \bibfield  {author} {\bibinfo {author} {\bibfnamefont {L.}~\bibnamefont
  {Einkemmer}}\ and\ \bibinfo {author} {\bibfnamefont {C.}~\bibnamefont
  {Lubich}},\ }\href {https://doi.org/10.1137/18M116383X} {\bibfield  {journal}
  {\bibinfo  {journal} {SIAM J. Sci. Comput.}\ }\textbf {\bibinfo {volume}
  {40}},\ \bibinfo {pages} {B1330} (\bibinfo {year} {2018})}\BibitemShut
  {NoStop}%
\bibitem [{\citenamefont {Klimas}\ and\ \citenamefont
  {Farrell}(1994)}]{klimas1994splitting}%
  \BibitemOpen
  \bibfield  {author} {\bibinfo {author} {\bibfnamefont {A.~J.}\ \bibnamefont
  {Klimas}}\ and\ \bibinfo {author} {\bibfnamefont {W.~M.}\ \bibnamefont
  {Farrell}},\ }\href {https://doi.org/10.1006/jcph.1994.1011} {\bibfield
  {journal} {\bibinfo  {journal} {J. Comput. Phys.}\ }\textbf {\bibinfo
  {volume} {110}},\ \bibinfo {pages} {150} (\bibinfo {year}
  {1994})}\BibitemShut {NoStop}%
\bibitem [{\citenamefont {Stanley}\ and\ \citenamefont
  {Ahlers}(1973)}]{stanley1973introduction}%
  \BibitemOpen
  \bibfield  {author} {\bibinfo {author} {\bibfnamefont {H.~E.}\ \bibnamefont
  {Stanley}}\ and\ \bibinfo {author} {\bibfnamefont {G.}~\bibnamefont
  {Ahlers}},\ }\href@noop {} {\emph {\bibinfo {title} {Introduction to phase
  transitions and critical phenomena}}}\ (\bibinfo {year} {1973})\BibitemShut
  {NoStop}%
\bibitem [{\citenamefont {Wu}\ \emph {et~al.}(2010)\citenamefont {Wu},
  \citenamefont {Lu}, \citenamefont {Huang},\ and\ \citenamefont
  {Wang}}]{wu2010transverse}%
  \BibitemOpen
  \bibfield  {author} {\bibinfo {author} {\bibfnamefont {M.}~\bibnamefont
  {Wu}}, \bibinfo {author} {\bibfnamefont {Q.}~\bibnamefont {Lu}}, \bibinfo
  {author} {\bibfnamefont {C.}~\bibnamefont {Huang}},\ and\ \bibinfo {author}
  {\bibfnamefont {S.}~\bibnamefont {Wang}},\ }\bibfield  {journal} {\bibinfo
  {journal} {J. Geophys. Res. Space}\ }\textbf {\bibinfo {volume} {115}},\
  \href {https://doi.org/10.1029/2009JA015235} {10.1029/2009JA015235} (\bibinfo
  {year} {2010})\BibitemShut {NoStop}%
\bibitem [{\citenamefont {Hutchinson}(2018)}]{hutchinson2018transverse}%
  \BibitemOpen
  \bibfield  {author} {\bibinfo {author} {\bibfnamefont {I.~H.}\ \bibnamefont
  {Hutchinson}},\ }\bibfield  {journal} {\bibinfo  {journal} {J. Plasma Phys.}\
  }\textbf {\bibinfo {volume} {84}},\ \href
  {https://doi.org/10.1017/S0022377818000909} {10.1017/S0022377818000909}
  (\bibinfo {year} {2018})\BibitemShut {NoStop}%
\bibitem [{\citenamefont {Hutchinson}(2019)}]{hutchinson2019transverse}%
  \BibitemOpen
  \bibfield  {author} {\bibinfo {author} {\bibfnamefont {I.~H.}\ \bibnamefont
  {Hutchinson}},\ }\href {https://doi.org/10.1103/PhysRevE.99.053209}
  {\bibfield  {journal} {\bibinfo  {journal} {Phys. Rev. E}\ }\textbf {\bibinfo
  {volume} {99}},\ \bibinfo {pages} {053209} (\bibinfo {year}
  {2019})}\BibitemShut {NoStop}%
\bibitem [{\citenamefont {Mangeney}\ \emph {et~al.}(2002)\citenamefont
  {Mangeney}, \citenamefont {Califano}, \citenamefont {Cavazzoni},\ and\
  \citenamefont {Travnicek}}]{mangeney2002numerical}%
  \BibitemOpen
  \bibfield  {author} {\bibinfo {author} {\bibfnamefont {A.}~\bibnamefont
  {Mangeney}}, \bibinfo {author} {\bibfnamefont {F.}~\bibnamefont {Califano}},
  \bibinfo {author} {\bibfnamefont {C.}~\bibnamefont {Cavazzoni}},\ and\
  \bibinfo {author} {\bibfnamefont {P.}~\bibnamefont {Travnicek}},\ }\href
  {https://doi.org/10.1006/jcph.2002.7071} {\bibfield  {journal} {\bibinfo
  {journal} {J. Comput. Phys.}\ }\textbf {\bibinfo {volume} {179}},\ \bibinfo
  {pages} {495} (\bibinfo {year} {2002})}\BibitemShut {NoStop}%
\bibitem [{\citenamefont {Colella}\ and\ \citenamefont
  {Woodward}(1984)}]{colella1984piecewise}%
  \BibitemOpen
  \bibfield  {author} {\bibinfo {author} {\bibfnamefont {P.}~\bibnamefont
  {Colella}}\ and\ \bibinfo {author} {\bibfnamefont {P.~R.}\ \bibnamefont
  {Woodward}},\ }\href {https://doi.org/10.1016/0021-9991(84)90143-8}
  {\bibfield  {journal} {\bibinfo  {journal} {J. Comput. Phys.}\ }\textbf
  {\bibinfo {volume} {54}},\ \bibinfo {pages} {174} (\bibinfo {year}
  {1984})}\BibitemShut {NoStop}%
\bibitem [{\citenamefont {Arber}\ and\ \citenamefont
  {Vann}(2002)}]{arber2002critical}%
  \BibitemOpen
  \bibfield  {author} {\bibinfo {author} {\bibfnamefont {T.}~\bibnamefont
  {Arber}}\ and\ \bibinfo {author} {\bibfnamefont {R.}~\bibnamefont {Vann}},\
  }\href {https://doi.org/10.1006/jcph.2002.7098} {\bibfield  {journal}
  {\bibinfo  {journal} {J. Comput. Phys.}\ }\textbf {\bibinfo {volume} {180}},\
  \bibinfo {pages} {339} (\bibinfo {year} {2002})}\BibitemShut {NoStop}%
\bibitem [{\citenamefont {Cheng}\ and\ \citenamefont
  {Knorr}(1976)}]{Cheng1976}%
  \BibitemOpen
  \bibfield  {author} {\bibinfo {author} {\bibfnamefont {C.~Z.}\ \bibnamefont
  {Cheng}}\ and\ \bibinfo {author} {\bibfnamefont {G.}~\bibnamefont {Knorr}},\
  }\href {https://doi.org/10.1016/0021-9991(76)90053-X} {\bibfield  {journal}
  {\bibinfo  {journal} {J.~Comput.~Phys.}\ }\textbf {\bibinfo {volume} {22}},\
  \bibinfo {pages} {330} (\bibinfo {year} {1976})}\BibitemShut {NoStop}%
\bibitem [{\citenamefont {Klimas}(1987)}]{klimas1987method}%
  \BibitemOpen
  \bibfield  {author} {\bibinfo {author} {\bibfnamefont {A.~J.}\ \bibnamefont
  {Klimas}},\ }\href {https://doi.org/10.1016/0021-9991(87)90052-0} {\bibfield
  {journal} {\bibinfo  {journal} {J. Comput. Phys.}\ }\textbf {\bibinfo
  {volume} {68}},\ \bibinfo {pages} {202} (\bibinfo {year} {1987})}\BibitemShut
  {NoStop}%
\bibitem [{\citenamefont {Klimas}\ \emph {et~al.}(2017)\citenamefont {Klimas},
  \citenamefont {Vi{\~n}as},\ and\ \citenamefont
  {Araneda}}]{klimas2017simulation}%
  \BibitemOpen
  \bibfield  {author} {\bibinfo {author} {\bibfnamefont {A.~J.}\ \bibnamefont
  {Klimas}}, \bibinfo {author} {\bibfnamefont {A.~F.}\ \bibnamefont
  {Vi{\~n}as}},\ and\ \bibinfo {author} {\bibfnamefont {J.~A.}\ \bibnamefont
  {Araneda}},\ }\bibfield  {journal} {\bibinfo  {journal} {J. Plasma Phys.}\
  }\textbf {\bibinfo {volume} {83}},\ \href
  {https://doi.org/10.1017/S002237781700054X} {10.1017/S002237781700054X}
  (\bibinfo {year} {2017})\BibitemShut {NoStop}%
\end{thebibliography}

%

\end{document}